\documentclass[usenatbib,usegraphicx,usepsfrag,onecolumn]{mn2e}
\usepackage[english]{babel} 
\usepackage{color}

\def\th{\vec{\theta}}
\def\u{\vec{U}}

\def\araa{ARAA}

\def\mnras{MNRAS}

\def\aap{A \& A}
\def\apj{Ap.J}

\def\apjl{Ap.JL}
\def\u{{\bf U}} 
\def\Sc{S_2}
\def\V2{V_2}
\def\V2ij{V_{2ij}}
\def\S{{\mathcal S}}
\def\V{\mathcal{V}}
\def\N{{\mathcal N}}

\def\la{\left\langle}

\def\lsim{~\rlap{$<$}{\lower 1.0ex\hbox{$\sim$}}}

\def\gsim{~\rlap{$>$}{\lower 1.0ex\hbox{$\sim$}}}

\usepackage{graphics}
\usepackage{color}
\usepackage{amsmath}
\usepackage{psfrag}
\begin{document}
\date {} 

\title[Visibility based angular power spectrum estimation] {Visibility based  angular power spectrum estimation in low frequency radio interferometric observations}
\author[S. Choudhuri et al.]{Samir Choudhuri$^{1}$\thanks{Email:samir11@phy.iitkgp.ernet.in}, Somnath Bharadwaj$^{1}$\thanks{Email:somnath@phy.iitkgp.ernet.in}, Abhik Ghosh$^{2}$ and SK. Saiyad Ali$^{3}$\\
  $^{1}$ Department of Physics,  \& Centre for Theoretical Studies, IIT Kharagpur,  Pin: 721 302, India\\
$^{2}$ Kapteyn Astronomical Institute, PO Box 800, 9700 AV Groningen, The Netherlands\\
$^{3}$ Department of Physics,Jadavpur University, Kolkata 700032, India}

\maketitle
   
\begin{abstract}
We present  two estimators  to quantify the angular 
power spectrum of the sky signal directly from the  visibilities measured
in radio interferometric observations.  This is relevant 
for  both the foregrounds and the cosmological $21$-cm 
signal buried therein. The discussion here is restricted to the 
Galactic synchrotron radiation, the most dominant 
foreground component after point source removal.   Our theoretical 
analysis is validated  using simulations at $150\, {\rm MHz}$, mainly for GMRT and also briefly for LOFAR.
The Bare Estimator uses  pairwise correlations of the measured 
visibilities, while the Tapered Gridded Estimator uses the visibilities 
after gridding  in the $uv$ plane. The former is very precise, but 
computationally  expensive for large  data. The latter 
has a lower precision, but takes less computation time which is proportional 
to the data volume. The latter also allows
tapering of the sky response leading to sidelobe suppression, an 
useful ingredient for foreground removal. Both  estimators   avoid 
the positive bias that arises due to  the system noise. 
We consider amplitude and phase errors of the gain,  and the $w$-term as possible sources of errors . We find that the estimated angular power spectrum  is exponentially sensitive
to the variance of the phase errors but insensitive to amplitude errors. 
The statistical uncertainties of the estimators are affected by both amplitude and 
phase errors. The $w$-term  does not have  a significant effect
at the angular scales of our interest. We propose the Tapered Gridded Estimator as an effective tool to observationally  quantify both foregrounds and  the  cosmological $21$-cm  signal.
\end{abstract} 


\section{Introduction}
Observations of the redshifted 21-cm radiation from the large scale
distribution of neutral hydrogen (HI) is one of the most promising
probes to study the high redshift Universe
(recent reviews: \citealt{morales10, mellema13}). This radiation appears as a
very faint, diffuse background radiation in all low frequency radio
observations below $1420 \,{\rm MHz}$. At these frequencies
the sky signal is largely dominated by different foregrounds which are four
to five orders of
magnitude stronger than the redshifted 21 cm signal
(\citealt{ali,bernardi09,ghosh150,pober}). Foreground removal 
is possibly the most serious challenge for detecting the  cosmological 
21-cm signal.  Various methodologies  have been explored  for 
foreground subtraction and  for detecting the underlying $21 \,{\rm cm}$
signal \citep{bowman,jelic,ghosh2,paciga11,petrovic,vedantham,mao3,liu2,cho,trott1,chapman1,chapman2,thyag,paciga13,jacobs,parsons14,dillon14,shaw14a,liu14a,liu14b}.

The Galactic synchrotron emission is expected
to be the most dominant foreground 
at angular scale $>10^{'}$ after point source subtraction at $10 -
20\, {\rm mJy}$ level
\citep{bernardi09,ghosh150}. A precise characterization and a detailed
understanding of the Galactic synchrotron  emission is needed to reliably
remove foregrounds in  $21\, {\rm cm}$ experiments.
The study of the Galactic synchrotron emission is interesting in its own
right.  This will  shed light on the cosmic ray electron distribution, 
the  strength and structure of the Galactic magnetic field,  and the
magnetic turbulence \citep{Waelkens,Lazarian,iacobelli13}. 

\cite{bernardi09} and \cite{ghosh150} have respectively analyzed 
$150 \, {\rm MHz}$
WSRT and GMRT observations where they find that the measured angular
power spectrum  can be well fitted with a power law 
($C_{\ell} \propto \ell^{-\beta}$, $\beta=2.2 \pm 0.3$ for WSRT 
and $\beta=2.34 \pm
0.28$ for GMRT) upto $\ell \le 900$. At relatively higher
frequencies, \citet{giardino01} and \citet{giardino02} have analyzed
the fluctuations in the Galactic synchrotron radiation using the $2.3
\, {\rm GHz}$ Rhodes Survey and the $2.4 \, {\rm GHz}$ Parkes radio
continuum and polarization survey, where they find a slope $\beta=2.43
\pm 0.01$ ($2 \le \ell \le 100$) and $\beta=2.37 \pm 0.21$ ($40 \le
\ell \le 250$) respectively. At tens of GHz, \cite{bennett03} have
determined the angular power spectrum of the Galactic synchrotron
radiation using the Wilkinson Microwave Anisotropy Probe (WMAP) data
where they find a scaling $C_{\ell} \sim \ell^{-2}$ within $\ell \le
200$. The structure of the Galactic synchrotron emission is not well
quantified  at the frequencies and angular scales relevant for detecting
 the 
cosmological $21$-cm signal, and there is considerable scope for further 
work in this direction.

Radio interferometric observations measure the complex visibility. 
The measurement is done directly in Fourier
space which makes interferometers ideal instruments for measuring the
angular power spectrum of the sky signal. 
The visibility based power spectrum
estimator formalism has been extensively used for analyzing CMB data
from interferometers (\citealt{hobson1,white,hobson2,myers}). 
A  visibility based estimator has also been successfully
employed to study the power spectrum of the HI in the 
interstellar medium (ISM) of several nearby  galaxies ( e.g. 
\citealt{Begum,  Dutta}).  A direct visibility based approach 
has been proposed for quantifying the power spectrum of the 
cosmological $21$-cm signal expected at the GMRT (\citealt{BS1,BP3,BA5})
and recently  for the ORT \citep{ali13}.  
Visibility based power spectrum estimators have been used to 
analyze   GMRT data in the context of HI observations
(\citealt{ali,paciga11,ghosh1,ghosh2,ghosh150}). A recent
paper  \citep{paul} has proposed  visibility correlations to detect  the EoR signal 
using  drift scan observations with the  MWA .

It is possible to  estimate the angular power spectrum of the 
sky signal from the synthesized radio image 
(e.g. \citealt{bernardi09,bernardi10,iacobelli13}). 
The noise properties of the visibilities are better understood
than those of the image pixels. The noise in the different 
visibilities is uncorrelated, whereas the noise in the image 
pixels may be correlated depending on the baseline $uv$ coverage. 
The visibility based power spectrum estimators have the added 
advantage that they avoid possible imaging artifacts due to the 
dirty beam, etc (\citealt{trott}).  

In this paper we consider two estimators which use the measured 
visibilities to quantify the angular power spectrum of the 
sky signal. The Bare Estimator, which has been utilized   in \citet{ali}
and \citet{ghosh1}, directly uses pairwise correlations of the measured
visibilities. 
The Tapered Gridded Estimator, which has been utilized  in 
\citet{ghosh2} and \citet{ghosh150}, uses the visibilities after gridding
on a rectangular grid in the $uv$ plane.  The latter incorporates 
the feature that it allows a tapering of the sky response and thereby 
suppresses the sidelobes of the telescope's primary beam. Earlier work 
\citep{ghosh2} has shown this to be a useful ingredient in foreground
removal for detecting the cosmological $21$-cm signal. In this paper
we have carried out a somewhat detailed  investigation 
in order to place these two estimators on sound theoretical footing. 
The theoretical predictions are substantiated using simulations. 
As a test-bed for the estimators, we consider a situation where 
the point sources have been identified and subtracted out so that 
the residual visibilities are dominated by the Galactic synchrotron 
radiation. We investigate how well the estimators are able to recover 
the angular power spectrum of the input model used to simulate
the Galactic synchrotron emission at $150 \, {\rm MHz}$. We have also analyzed the effects
of gain errors and the $w$-term. Most of our simulations are for the 
GMRT, but we also briefly consider simulations for  LOFAR. 
The estimators considered here can be generalized to the multi-frequency 
angular power spectrum  (MAPS, \citealt{kanan})  which can be used 
to quantify the cosmological $21$-cm signal. We plan to investigate this
 in a future study. 

A brief outline of the paper follows. In Section 2 we  establish the
relation between the visibility correlation and the angular power spectrum. 
In Section 3 we describe the simulations which we have used to validate 
the theoretical results of this paper.  In Sections 4 and 5 
we consider the Bare and the Tapered Gridded Estimators respectively. 
The theoretical analysis and the results from the simulations are 
all presented in these two sections. 
Section 6 presents a brief comparison between the two estimators,
and in Sections 7 and 8 we consider the effect of gain errors and the 
$w$-term respectively. Much of the analysis of the previous sections
is in the context of the GMRT. In Section 9 we apply the estimators
to simulated LOFAR data and present the results.  We present discussion 
 and conclusions in Section 10. 

\section{Visibility Correlations and the angular power spectrum}
\label{v2ps}
In this section we discuss  the  relation between the two visibility 
correlation and the  angular power spectrum of the specific intensity
$I(\th,\,\nu)$ or equivalently the brightness temperature $T(\th,\,\nu)$
distribution  on the sky   under the flat-sky approximation.
 Here $\th$ is a two   dimensional vector on the
plane of the sky with origin at the center of the field of view
(FoV). It is useful to decompose the specific intensity 
as $ I(\th,\nu)=\bar{I}(\nu)+\delta I(\th,\,\nu)$ where the first
term $\bar{I}(\nu) $ is an uniform background brightness and the
second term $\delta I(\th,\,\nu)$ is the angular fluctuation in the
specific intensity. We assume that  $\delta I(\th,\,\nu) $ is a
particular realization of a statistically homogeneous and isotropic
Gaussian random process on the sky. In radio interferometric observations, the
fundamental observable quantity is a set of complex visibilities
$\V(\u,\nu)$ which are sensitive to only the angular fluctuations in
the sky signal. The baseline $\u$ quantifies the antenna pair separation
${\bf d}$ projected on the plane perpendicular to the line of sight in
units of the observing wavelength $\lambda$. The 
 measured visibilities   are a sum of  two contributions 
$\V(\u, \nu)=\S(\u, \nu)+\N(\u,\nu)$,
the  sky signal and system noise respectively.
We assume 
that the signal and the noise are both uncorrelated Gaussian random 
variables with zero mean.  The
visibility contribution $\S(\u,\nu)$ from  the sky signal records the
Fourier transform of the product of the primary beam pattern
${\mathcal A}(\th, \nu)$ and $\delta I(\th,\,\nu)$. The primary beam pattern
${\cal A}(\th, \nu)$ quantifies how the individual antenna responds to
signals from different directions in the sky.  Using the convolution
theorem, we then have
\begin{equation}
\S(\u,\nu)=  \int \, d^2 U{'}  \,
  \tilde{a}\left(\u - \u{'},\,\nu\right)\, 
 \, \Delta \tilde{I}(\u{'},\,\nu),   
\label{eq:a2}
\end{equation}
where $\Delta \tilde{ I}(\u,\,\nu)$ and $\tilde{a}\,(\u, \nu)$ are the
Fourier transforms of $\delta I(\th,\,\nu) $ and ${\cal A}(\th,\,\nu)$
respectively. Typically, the term arising from the uniform specific 
intensity distribution 
 $\bar{I}(\nu) \tilde{a}\,(\u, \nu)$ makes no contribution to
the measured visibilities, and we have dropped this.
 We refer to $\tilde{a}\,(\u, \nu)$ as the
aperture power pattern.  The individual antenna response ${\cal A}(\th, \nu)$ 
 for any telescope is usually quite complicated
depending on  the  telescope aperture, the reflector and 
the feed {\citep{lfra}}. It is beyond
the scope of the present paper to consider the actual single antenna 
response of any particular telescope. We make the simplifying assumption
that the telescope has an uniformly illuminated circular aperture 
of diameter $D$ 
whereby we have the primary beam pattern (Figure 1)
\begin{equation}
{\cal A}(\th,\,\nu) = 
\left[ \left(\frac{2 \lambda}{\pi\theta D} \right)
J_1\left(\frac{\pi\theta D}{\lambda}\right) \right]^2 
\label{eq:b1} 
\end{equation}
 where $J_1$ is the Bessel function of the first kind of order one, the
 primary  beam pattern  is normalized to unity  at the pointing center
$[{\cal A}(0)=1]$, 
and  the aperture power pattern is  
\begin{equation}
\tilde{a}(\u, \nu)=\frac{8 \lambda^4}{\pi^2 D^4}\bigg[\bigg(\frac{D}{\lambda}\bigg)^2 \cos^{-1}\bigg(\frac{\lambda U}{D}\bigg)-U\sqrt{\bigg(\frac{D}{\lambda}\bigg)^2-U^2}\bigg],
\label{eq:b2} 
\end{equation}

We note that  $\tilde{a}(\u, \nu)$ in  eq. (\ref{eq:b2}) peaks at $U=0$,
declines monotonically with increasing $U$,  and is zero for 
 $U \ge  D/{\lambda}$. The primary beam pattern (Figure 1) is 
well approximated by a circular Gaussian function 
\begin{equation}
{\cal A}_G(\th,\nu)=\exp[-\theta^2/\theta^2_0]
\label{eq:b1a} 
\end{equation} 
of  the same full width at half maxima (FWHM) as eq. (\ref{eq:b1}). 
The parameter $\theta_{0}$ here   is related to the
 full width half maxima  $\theta_{\rm FWHM}$ of the primary beam pattern
 ${\cal A}(\th,\nu)$ (eq. \ref{eq:b1})  as ${\theta}_{0} =0.6 \theta_{\rm FWHM}$,  
and 
\begin{equation}
\tilde{a}_G(\u, \nu)=\frac{1}{\pi U_0^2} \ e^{-U^2/U_0^2}
\label{eq:b2a} 
\end{equation}
where $U_0=(\pi \theta_0)^{-1}=0.53/ \theta_{\rm FWHM} $.  While the Gaussian
$\tilde{a}_G(\u, \nu)$  (eq. \ref{eq:b2a}) provides a  good
approximation  to $\tilde{a}(\u, \nu)$ (eq. \ref{eq:b2}) particularly 
in the vicinity of $U=0$, there is  however  a significant difference in 
that $\tilde{a}(\u, \nu)$ has a compact support and is exactly zero for all 
$U \ge  D/{\lambda}$ whereas $\tilde{a}_G(\u, \nu)$, though it has an 
extremely small value for $U \ge  D/{\lambda}$,  does not become 
zero anywhere.  In practice 
it is extremely difficult to experimentally determine the full primary 
beam pattern ${\cal A}(\th,\,\nu)$ for a telescope. However,  
the value of $\theta_{\rm FWHM}$ is typically well determined. 
This has motivated the Gaussian approximation to be used extensively 
for both theoretical predictions \citep{BS1,BA5} and analyzing observational 
data \citep{ali,ghosh150}. 
The close
match between ${\cal A}(\th,\,\nu)$ (eq. \ref{eq:b1})  and 
${\cal A}_G(\th,\,\nu)$ (eq. \ref{eq:b1a}) indicates that we
 may also expect the 
Gaussian approximation to provide a good fit 
to the telescope's actual primary beam pattern, particularly within the main
lobe.  This, to some extent, justifies the use of the Gaussian approximation
 in the earlier works.  The Gaussian approximation  simplifies the calculations
rendering them amenable to analytic treatment, and we use it on several 
occasions as indicated later in this paper.  For much of the investigations
presented in this paper we have considered $D=45 \ {\rm m}$ and 
$\lambda=2 \ {\rm m}$  which corresponds to    GMRT  $150 \ {\rm MHz}$ 
observations. We have also considered $D=30.75  \ {\rm m}$ and  
$\lambda=2 \ {\rm m}$ which corresponds to  LOFAR $150 \ {\rm MHz}$ 
observations.  For both these telescopes, 
Table~\ref{tab:1} summarizes  the values of some of the relevant parameters. 
Note that these values correspond to the idealized telescope model
discussed above, and they are somewhat different from the values 
actually measured for the respective telescopes. For example, the GMRT
primary beam pattern has $\theta_{\rm FWHM}=186^{'}$ whereas we have used
$\theta_{\rm FWHM}=157^{'}$ based on  our idealized  model.  We discuss 
the observational consequence of this $\sim 16\%$ difference later in 
 Section~\ref{sum} of this paper. For the rest of this paper we focus on the 
GMRT , except in  Section~\ref{lofar} where we shift our attention to LOFAR. 
Our entire analysis is based on the idealized telescope model described
above and  the relevant  parameters are listed in 
Table~\ref{tab:1}  for both these telescopes.

\begin{table}
\begin{center}
\begin{tabular}{|l|cc|c|c|c|c|}
\hline
$150 \, {\rm MHz}$&$D$  &$\theta_{\rm FWHM}$ & $\theta_0$ & $U_0$ & $\sigma_0$ \\
\hline
& &$1.03  \lambda/D$  & $0.6  \theta_{\rm FWHM} $ & $0.53/ \theta_{\rm FWHM} $ & $0.76/  \theta_{\rm FWHM} $\\
\hline
GMRT& $45 \, {\rm m}$  & $157^{'}$& $95^{'}$ &11.54 & 16.6\\
\hline
LOFAR & $30.75\, {\rm m}$ & $230^{'}$& $139^{'}$& 7.88& 11.33\\
\hline
\end{tabular} 
\caption{This shows some  relevant parameters for the primary beam pattern calculated using the 
idealized telescope model (eqs. \ref{eq:b1},\ref{eq:b2}),
and the Gaussian approximation  (eqs. \ref{eq:b1a},\ref{eq:b2a}). The parameter $\sigma_0$ is defined
in eq.~(\ref{eq:dellu}).}  
\label{tab:1}
\end{center}
\end{table}

\begin{figure}
\begin{center}
\psfrag{theta}[c][c][1.5][0]{$\theta$ $[{\rm arcmin}$]}
\psfrag{Pbeam}[c][c][1.5][0]{${\mathcal A}(\theta,\nu)$}
\psfrag{Bessel}[r][r][1.3][0]{Model}
\psfrag{Gaussian}[r][r][1.3][0]{Gaussian}
\includegraphics[width=70mm,angle=-90]{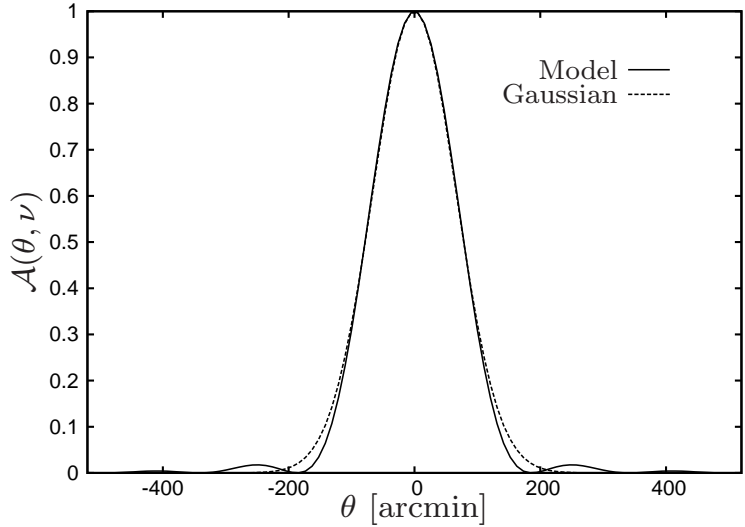}
\caption{The solid curve shows the $150\, {\rm  MHz}$   GMRT primary beam pattern ${\cal A}(\th,\,\nu)$ 
predicted by eq. (\ref{eq:b1}), and the dashed curve shows Gaussian approximation  (eq. \ref{eq:b1a}) with 
the same $\theta_{\rm FWHM}$.}
\label{fig:gauss}
\end{center}
\end{figure}

In the flat sky approximation 
the statistical properties of the background intensity fluctuations
$\delta I(\th,\,\nu)$ can be quantified through the two dimensional (2D)
 power spectrum  $P(U, \nu)$ defined as,
\begin{equation}
 \langle \Delta \tilde{I}(\u ,\nu) \Delta \tilde{I}^{*}(\u' ,\nu) \rangle = 
\delta_{D}^{2}(\u-\u') \, P(U, \nu),
\label{eq:v4}
\end{equation}where $\delta_{D}^{2}(\u-\u')$ is a two dimensional Dirac delta
function. The angular brackets $\langle ... \rangle$ here denote an
ensemble average over different realizations of the stochastic
intensity fluctuations on the sky. We also assume that the $P(U,
\nu)$ depends only on the magnitude $U=|\u|$  {\it i.e.} the fluctuations
are statistically isotropic. 
We note that $ P(U, \nu)$  is related to $C_{\ell}(\nu)$ the
angular power spectrum of the brightness temperature fluctuations
 through \citep{ali} 
\begin{equation}
C_{\ell}(\nu) =\left( \frac{\partial B}{\partial T}\right)^{-2}
P(\ell/2 \pi, \nu) \,,
\label{eq:c2}
\end{equation}  
where the angular multipole $\ell$ corresponds to 
$U=\ell/2 \pi$, $B$ is the Planck function and $({\partial B}/{\partial T})=2
k_B/\lambda^2$ in the Raleigh-Jeans limit which is valid at the
frequencies of our interest. We will drop the $\nu$ dependence henceforth as 
the rest of the calculations are done at a fixed  frequency $ \nu = 150\, 
{\rm  MHz}$.

We now consider the two visibility correlation which is defined as 
\begin{equation}
V_2(\u, \u+\Delta \u)=\langle  \V(\u){\V}^*(\u+\Delta \u) \rangle \,,
\end{equation}
and which has the contribution 
\begin{equation}
S_2(\u, \u+\Delta \u) = \int d^2 U{'} \, \tilde{a}(\u-\u{'}) \, \tilde{a}^{*}(\u+\Delta \u-\u{'})\,P(U{'}) \,
\label{eq:v6}
\end{equation}
from the sky signal. 

\begin{figure}
\begin{center}
\psfrag{1000Para}[r][r][1][0]{$\u=1,000$}
\psfrag{alpha}[Br][r][1][0]{$\beta=2.34$}
\psfrag{gaussian}[br][r][1.2][0]{Gaussian}
\psfrag{x}[t][c][1][0]{$\Delta$\u}
\psfrag{y}[b][c][1][0]{$S_2(\u, \u+\Delta \u)/S_2(U)$}
\includegraphics[width=100mm,angle=0]{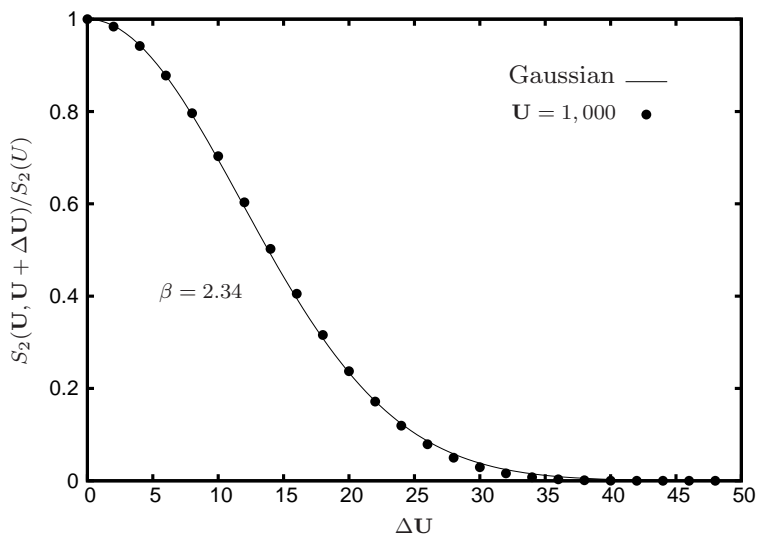}
\caption{This figure shows how the sky signal contribution to the two visibility correlation 
  varies with $\Delta\u$ for a fixed  value $U=1,000$. The points  show the 
results from eq. (\ref{eq:v6}) for $P(U)=A U^{-2.34}$,
and the solid line shows the  Gaussian fit  given in eq.  (\ref{eq:dellu}). }
\label{fig:deltau}
\end{center}
\end{figure}

The visibilities at the baselines $\u$ and $\u+\Delta\u$ are  correlated
 only if there is a significant overlap between 
 $\tilde{a}(\u-\u')$ and $\tilde{a}^{*}(\u+\Delta \u-\u')$.
The correlation  $S_2(\u,\u+\Delta\u)$  is strongest when $|\Delta \u| =0$,
declines rapidly with increasing $|\Delta \u|$,  and is zero for 
 $|\Delta \u| \ge 2D/\lambda$.  The correlation  $S_2(\u,\u+\Delta\u)$ 
depends on both, the magnitude of 
$\Delta \u$ as well as the angle between $\Delta \u$ and $\u$, and an 
earlier work \citep{BP3} has studied this in detail for the predicted 
post-reionization cosmological 21-cm signal.  In this work we have 
considered a power law power spectrum  $P(U)=A U^{-\beta}$ 
for different values of $\beta$ in the range $1.5$ to $3.5$,
and  we have used eq.~(\ref{eq:v6}) to study the  $\Delta \u$ dependence of 
$S_2(\u,\u+\Delta\u)$. We find that the  $\Delta\u$ dependence is isotropic 
to a great extent, and it can be well modelled using a Gaussian 
(Figure \ref{fig:deltau}) as 
\begin{equation} 
S_2(\u, \u+\Delta \u) = 
\exp\bigg[-\bigg(\frac{\mid \Delta \u \mid }{\sigma_0}\bigg)^2\bigg]
\, S_2(U),
\label{eq:dellu}
\end{equation}
where $\sigma_0=0.76/\theta_{\rm FWHM}$ (Table~\ref{tab:1}) and $S_2(U) \equiv 
S_2(\u,\u)$.  While the approximation in eq.~(\ref{eq:dellu}) matches the result of 
 eq.~(\ref{eq:v6}) quite well for small $\Delta \u$,  the approximation breaks down 
when  $\mid \Delta  \u \mid > 2 D/\lambda$ where $S_2(\u, \u+\Delta \u)=0$ contrary 
to the prediction of  eq.~(\ref{eq:dellu}). This discrepancy, however, does not significantly 
affect the estimators (defined later) because the value of   $S_2(\u, \u+\Delta \u)$ 
predicted by eq.~(\ref{eq:dellu})  is extremely small for  $\mid \Delta  \u \mid > 2 D/\lambda$.

\begin{figure}
\begin{center}
\psfrag{cl}[b][t][1][0]{$S_2(U)$}
\psfrag{U}[r][cB][1][0]{U}
\psfrag{a2.3}[r][c][1][0]{$\beta=2.34$}
\psfrag{a1.8}[l][c][1][0]{$\beta=1.8$}
\includegraphics[width=100mm,angle=0]{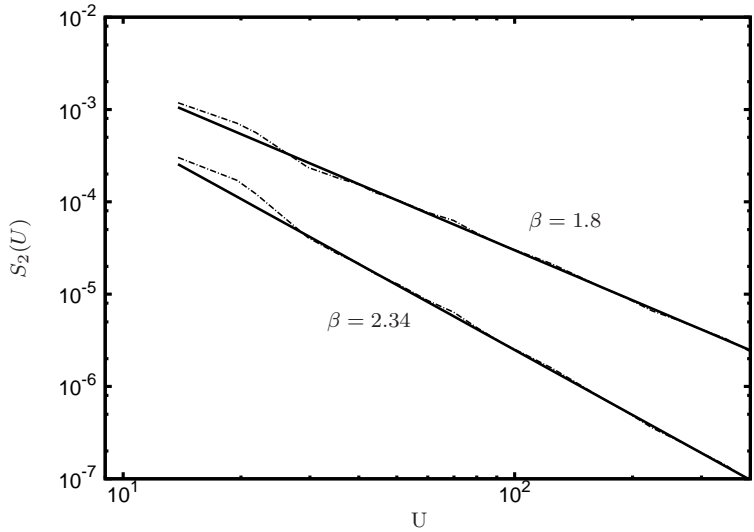}
\caption{This  shows the sky signal contribution to the visibility correlation ($S_2(U)$) 
for  two different power spectra with  slopes $\beta=1.8$ and $2.34$ respectively. 
The dash-dot curve shows the result of the convolution  in  eq.(\ref{eq:v6}) 
with $\Delta \u=0$   whereas the solid curve shows the result of  approximating this 
with eq. (\ref{eq:v6a}). We see that the approximation of eq. (\ref{eq:v6a})  matches 
the convolution reasonable well at large baselines $U \ge 4 U_0 \sim 45$.}
\label{fig:psconv}
\end{center}
\end{figure}

A further simplification is possible  for $U \gg U_0$ where 
it is possible to approximate  $S_2(U)$ which is calculated using  eq.~(\ref{eq:v6}) by  assuming that 
the value of $P(U{'})$ does not  change much within 
the width of the function $|\tilde{a}\left(\u -  {\u}{'} \right)|^2 $. 
We then obtain  
\begin{equation}
S_2(U) = \left[\int d^2 U{'} \, \mid \tilde{a}(\u-\u{'}) \mid^2 \right]
P(U) \,.    
\label{eq:v6a}
\end{equation}
The integral in the square brackets has a constant value 
$\frac{\pi\theta_0^2}{2}\,$ in the Gaussian approximation which  yields the 
value $1.19\times10^{-3}$,  whereas we have $1.15\times10^{-3}$
if we use eq.~(\ref{eq:b2}) and numerically evaluate the integral in the 
square brackets. We see that the  Gaussian approximation is adequate for 
 the integral in  eq.~(\ref{eq:v6a}), and we adopt the value $\pi \theta_0^2/2$ 
for the entire subsequent analysis. 
We have calculated  $S_2(U)$  (Figure.\ref{fig:psconv})
using the convolution  in eq.~(\ref{eq:v6}), and compared this with the  
approximation in eq.~(\ref{eq:v6a}). We find that the approximation 
in eq.~(\ref{eq:v6a})  matches quite well with the convolution  (eq. \ref{eq:v6})
for baselines $U \ge 4 U_0 \sim 45$. Throughout  the subsequent analysis we have restricted 
the baselines to this range, and  we have used  eq.~(\ref{eq:v6a}) to evaluate  $S_2(U)$,
the  sky signal contribution to the visibility correlation.

We finally have the approximate relation between the sky signal contribution 
to the two visibility 
correlation and the angular power spectrum 
\begin{equation}
\Sc(\u, \u+\Delta \u) = \frac{\pi \theta_0^2}{2} 
\left( \frac{\partial B}{\partial T}\right)^{2}
\exp\bigg[-\bigg(\frac{\Delta U}{\sigma_0}\bigg)^2\bigg] 
\, C_{\ell}
\label{eq:vcf}
\end{equation}
where $\ell= 2 \pi U$.  We thus see that the visibilities at two different
baselines $\u$ and $\u + \Delta \u$ are correlated only if the separation
is small $(\mid \Delta U \mid \le \sigma_0)$, and there is negligible correlation
if the separation is beyond a disk of radius $\sigma_0$. Further, the visibility
correlation $S_2(\u, \u+\Delta \u)$  
gives a direct estimate of the angular power spectrum $C_{\ell}$
at the angular 
multipole $\ell = 2 \pi U$. In addition to the sky signal $\S(\u)$, 
each  visibility also contains  a system noise contribution $\N(\u)$. 
For each visibility measurement, the real and imaginary parts of $\N(\u)$  are both   
random variables of zero mean and rms. $\sigma_n$. Further, the noise
in any two different visibilities is uncorrelated. We can then write the total
visibility correlation as 
\begin{equation}
V_{2ij} \equiv  \langle \V_i \V^{*}_j \rangle = V_0 \, e^{-\mid \Delta \u_{ij} \mid^2/
\sigma_0^2} 
\, C_{\ell_i} + \delta_{ij} 2 \sigma_n^2
\label{eq:vcorr}
\end{equation}
where $ [\V_i,\V_j] \equiv [\V(\u_i),\V(\u_j)]$, $V_0= \frac{\pi \theta_0^2}{2}
\left( \frac{\partial B}{\partial T}\right)^{2}$,  $\Delta \u_{ij}=
\u_{i}-\u_{j}$ and the Kronecker delta $\delta_{ij}$ is nonzero only if 
we correlate a visibility with itself. Equation (\ref{eq:vcorr}) relates
the two visibility correlation $V_{2ij}$ to $C_{\ell_i}$ the angular power 
spectrum  of the sky signal at the angular multipole  $\ell_i=2 \pi U_i$ 
and $\sigma_n^2$ the mean square system  noise,  
and we use this extensively in connection with 
 the estimators that we consider in the subsequent sections.

\section{Simulating the sky signal}
\label{ps_simu}
We have used simulations of radio-interferometric observations to validate
the angular power spectrum estimators that we introduce in subsequent
sections of this paper.  In this section  we first describe  the simulations    
of the sky signal,  and  then describe how these were used to simulate
the expected visibilities. For the sky model, we assume that all 
point sources 
with flux above a sufficiently low threshold have been identified and removed
from the data so that the $150 \, {\rm MHz}$ radio sky is dominated 
by the diffuse Galactic Synchrotron radiation. 

The slope $\beta$ of  the angular power spectrum of diffuse Galactic synchrotron emission  is within the
 range $1.5$ to $3$ as found by all the previous measurements at frequencies $0.15 -94 \, {\rm GHz}$ 
(e.g. \citealt{laporta,bernardi09}). 
For the purpose of this paper we assume that the 
fluctuations in the 
diffuse Galactic Synchrotron radiation are a statistically homogeneous
and isotropic Gaussian random field  whose  statistical properties 
are completely specified by the  angular power spectrum. Further, we
assume that the angular power spectrum of brightness temperature 
fluctuations is well described by a single 
power law  over the entire range of angular scales of our interest. 
In this work we have adapted the angular power spectrum 
\begin{equation}
C^M_{\ell}=A_{\rm 150}  \times \left(\frac{1000}{\ell} \right)^{\beta},
\label{eq:cl150}
\end{equation}
where  $A_{\rm 150}=513 \, {\rm mK}^2$ and $\beta=2.34 $.
from \citet{ghosh150}. This is the input model for all our simulations.  

\begin{figure}
\begin{center}
\includegraphics[width=120mm,angle=0]{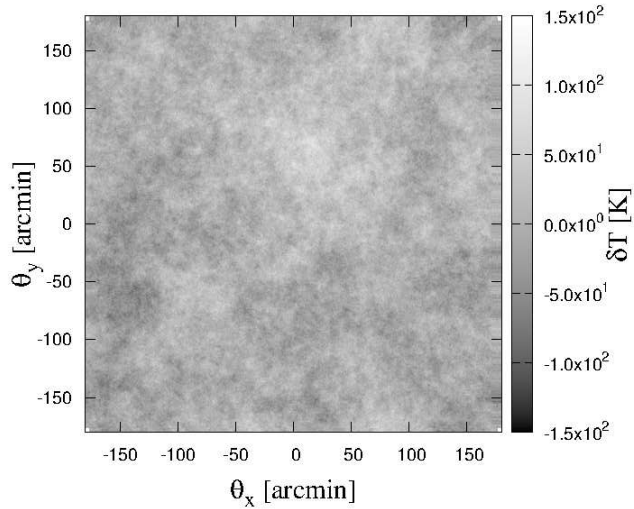}
\caption{This shows a single realization of the simulated $150 \, {\rm MHz}$ radio sky 
under the assumption that the bright point sources have been removed so that it is 
dominated by the diffuse Galactic synchrotron radiation. We have simulated a 
$5.8^{\circ} \times 5.8^{\circ}$ FoV with $\sim 10.2''$ resolution. }
\label{fig:grf}
\end{center}
\end{figure}

\begin{figure}
\begin{center}
\includegraphics[width=110mm,angle=0]{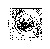}
\caption{This shows the $uv$  coverage  for $8 \, {\rm hr}$  GMRT $150 \, {\rm
    MHz}$  observations centered  on a field at a declination of $\delta=+60^{\circ}$. 
Only baselines with $\mid u \mid, \mid v \mid \le 1,000$ have been shown. 
Note that $u$ and $v$ are antenna separations measured in units of the observing wavelength,
and hence they are dimensionless.}
\label{fig:gmrtbasl}
\end{center}
\end{figure}

We have considered a  $5.8^\circ \times 5.8^\circ$  FoV  
for the GMRT simulations. This has been 
represented using  a $2048\times2048$ grid  with an angular resolution  of 
$\sim10.2$ arcsecond .  We have first generated the  Fourier components of 
the brightness temperature fluctuations on the grid using ,
\begin{equation}
\Delta \tilde{T}(\u)=\sqrt{\frac{\Omega \, C^M_{\ell}}{2}}[x(\u)+iy(\u)],
\label{eq:ran}
\end{equation}
where $\Omega$ is the total solid angle of the simulation, and 
$x(\u)$  and $y(\u)$   are  independent Gaussian random variables with 
zero mean and unit variance.  We then use  a Fourier transform
to generate the brightness temperature fluctuations  $\delta T(\th)$ or 
equivalently the specific intensity fluctuations $\delta I(\th)$
on the grid.   Figure~\ref{fig:grf} shows one  realization
of the brightness temperature  fluctuations generated using the procedure
outlined above. We have generated $20$ different independent realizations 
of the sky by considering different sets of random numbers in 
eq.~(\ref{eq:ran}).

To simulate GMRT observations we consider $8 \, {\rm hr}$  observations 
targeted on a field located at $+60^{\circ}$ DEC  for which the $uv$ tracks 
for baselines within $\mid u \mid, \mid v \mid  \le 1,000$ are shown in
 Figure \ref{fig:gmrtbasl}.
We assume $16 {\rm s}$  integration time for each sampled visibility data
which gives us  $2,17,457$ visibility points. To calculate 
the visibilities, we have multiplied the simulated $\delta I(\th)$  
with the primary beam pattern ${\mathcal A}(\th)$ (eq.~\ref{eq:b1})
and evaluated  the Fourier transform  of the product for each sampled
baseline $\u$ on the $uv$ track. In addition to the sky signal, each 
measured visibility will also have a system noise contribution. We have 
included this by adding  independent Gaussian  random noise  contributions
to both the real and imaginary parts of each visibility. This noise
is predicted to have an rms. of $\sigma_n=1.03 \, {\rm Jy}$ for
a single polarization at  the GMRT\footnote{http://www.gmrt.ncra.tifr.res.in}.  

It is clearly visible in Figure \ref{fig:gmrtbasl} that the GMRT has a 
rather  sparse $uv$ coverage. The fact that we  have data for only a  
limited number  of the Fourier modes is expected to play an important role.
 This is  particularly  important for the cosmic variance 
which crucially depends on the number of independent Fourier modes. 
 In order to assess the impact  of the  sparse $uv$ 
coverage we have also considered a situation where exactly the same 
number of
visibility measurements  ($2,17,457$) are randomly  distributed 
within the region $\mid u \mid, \mid v \mid  \le 1,000$
on the  $uv$ plane.  

In the subsequent sections of this paper we have analyzed $20$ independent 
realizations of the sky signal,  with visibilities points that 
correspond to the $uv$ tracks shown in  Figure \ref{fig:gmrtbasl}. We 
refer to this ensemble of 20 simulated data sets 
as ``GMRT''.  We have also considered a random baseline
distribution and calculated the visibilities for the same $20$ realizations of the sky signal, 
and we  refer to this as ``Random''.  Finally, we have also carried out simulations 
for LOFAR which has a more uniform $uv$ coverage as compared to the GMRT.
These simulations are separately discussed in Section \ref{lofar}.

Finally, we note that the simulated baselines lying in the lower half of the $uv$ plane 
(e.g.  Figure \ref{fig:gmrtbasl}.)  are all folded  to the upper half  using the property 
$\V(\u)=\V^{*}(-\u)$.  The simulated baseline distribution that  we finally use for analysis 
is entirely restricted to the upper half of the $uv$ plane.

%

\section{The Bare Estimator}
\label{sec:bare}
The Bare Estimator  directly uses  the individual visibilities to estimate the 
angular power spectrum. Each measured visibility corresponds to a Fourier mode of the 
sky signal, and  the visibility squared  $\mid \V \V^{*} \mid$  straight away gives the angular power  
spectrum. This  simple estimator, however, has a severe drawback because the noise contribution 
$ 2 \sigma^2_n$ is usually much larger than the  sky signal 
$ V_0 \, e^{-\mid \Delta \u_{ij} \mid^2/ \sigma_0^2} \, C_{\ell}$ in eq.~(\ref{eq:vcorr}). 
Any estimator that  includes the correlation of a visibility with itself suffers from a very large 
positive noise bias. It is, in principle, possible to model the constant noise bias and subtract it out.
This however   is extremely difficult in practice because  small calibration errors 
(discussed later in Section~\ref{sec:gerr})  would  introduce fluctuations in the noise bias 
 resulting  in  residuals that could exceed the sky signal. It is therefore desirable to avoid the 
noise bias by considering estimators which  do not include  the  contribution from the correlation of a
visibility with itself. 

The   Bare Estimator $\hat E_B(a)$ is defined as 
\begin{equation}
\hat E_B(a)=\frac{\sum_{i,j} \, w_{ij} \, \V_{i} \,  \V^{*}_{j} }{\sum_{i,j} w_{ij} V_0  
e^{-\mid \Delta \u_{ij} \mid^2/\sigma_0^2} } \,, 
\label{eq:be1}
\end{equation}
where we have  assumed that the  baselines have been divided into bins such that all the baselines
$U$ in the range $U_1 \le U < U_2$ are in  bin $1$, those in the range $U_2 \le U < U_3$ 
are in  bin $2$  etc., and $\hat E_B(a)$ refers to  a particular  bin  $a$.  The sum $i,j$ is  over all pairs of 
visibilities $\V_i,\V_j$  with  baselines $\u_i,\u_j$  in  bin $a$. 
We have restricted the sum to pairs within $\mid \u_i - \u_j \mid \le \sigma_0$
as the pairs with larger separations do not contribute much to the estimator. 
 The weight 
$w_{ij}=(1-\delta_{ij}) K_{ij}$ is chosen such that 
it is zero when we correlate a visibility with itself, thereby avoiding the positive noise bias. 

We now show that  $\hat E_B(a)$ gives an unbiased estimate of the angular 
power spectrum $C_{\ell}$ for bin $a$. 
The expectation value of the estimator can be expressed using eq. (\ref{eq:vcorr}) as 
\begin{equation}
\langle \hat E_B(a) \rangle 
=\frac{\sum_{i,j}  \, w_{ij} \, V_{2ij} }{\sum_{i,j}  w_{ij} V_0  
e^{-\mid \Delta \u_{ij} \mid^2/\sigma_0^2} }  
=\frac{\sum_{i,j} \, w_{ij} \,e^{-\mid \Delta \u_{ij} \mid^2/\sigma_0^2} C_{\ell_i}  }{\sum_{i,j}  w_{ij}  
e^{-\mid \Delta \u_{ij} \mid^2/\sigma_0^2} }  
\label{eq:be2}
\end{equation}
which can be written as 
\begin{equation}
\langle \hat E_B(a) \rangle = \bar{C}_{\bar{\ell}_a} 
\label{eq:be3}
\end{equation}
where $ \bar{C}_{\bar{\ell}_a}$ is the average  angular power spectrum  at 
 \begin{equation}
\bar{\ell}_a
=\frac{\sum_{i,j} \, w_{ij} \,e^{-\mid \Delta \u_{ij} \mid^2/\sigma_0^2} \ell_i  }{\sum_{i,j}  w_{ij}  
e^{-\mid \Delta \u_{ij} \mid^2/\sigma_0^2} }  \,.
\label{eq:be4}
\end{equation}
which is the   effective angular multipole  for bin $a$. 

We note that it is possible to express eq.~(\ref{eq:be2})  using matrix notation 
as
\begin{equation}
\langle \hat E_B(a) \rangle 
=\frac{Tr({\bf w} {\bf V}_2)}{Tr({\bf w} {\bf I}_2)}
\label{eq:be4b}
\end{equation}
where we have the matrices ${\bf w} \equiv w_{ij}$, ${\bf V}_2 \equiv V_{2ij}$, ${\bf I}_2= V_0  
e^{-\mid \Delta \u_{ij} \mid^2/\sigma_0^2}$ and $Tr({\bf A})$ denotes the trace of a matrix ${\bf A}$. 

We next evaluate  $\sigma^2_{E_B}(a)$  the variance of $\hat E_B(a)$. This  gives
$\delta C_{\ell_a}$  which is 
an estimate of the error in the  angular power spectrum measured from the data. 
We have 
\begin{equation}
[\delta C_{\ell_a}]^2 \equiv \sigma^2_{E_B}(a)=\langle \hat E^2_B(a) 
\rangle - \langle \hat E_B(a) \rangle^2
\label{eq:be5}
\end{equation} 
which can be simplified to 
\begin{equation}
\sigma^2_{E_B}(a)=\frac{\sum_{i,j,k,l} w_{ij} w_{kl} V_{2il} V_{2kj}}{[Tr({\bf w} 
{\bf I}_2)]^2}
=\frac{Tr({\bf w} {\bf V}_2 {\bf w} {\bf V}_2)}{[Tr({\bf w} {\bf I}_2)]^2}
\label{eq:be6}
\end{equation} 
under the assumptions  that  ${\bf w}$ is symmetric and  the measured 
visibilities are Gaussian 
random variables.

\begin{figure}
\begin{center}
\psfrag{cl}[b][t][1.5][0]{$\ell (\ell+1) C_{\ell}/2 \pi \, [mK^2]$}
\psfrag{U}[c][c][1.5][0]{$\ell$}
\psfrag{Ghosh}[cr][tr][1.2][0]{$C^M_{\ell}$}
\psfrag{gmrt}[r][r][1.2][0]{GMRT}
\includegraphics[width=100mm,angle=0]{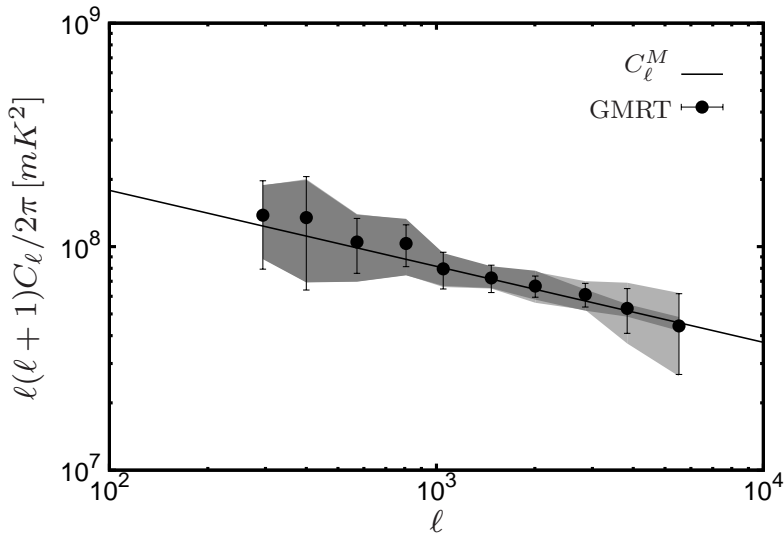}
\caption{This shows $C_{\ell}$ multiplied with $\ell (\ell + 1)/2 \pi$, 
plotted as a function of $\ell$. 
The  solid  line shows the input model (eq.~\ref{eq:cl150}) used for the 
simulations, and the points show the values recovered  by the  Bare
Estimator (eq.~\ref{eq:be1}).  The points show the mean and  the light  shaded 
region shows the $1\sigma$ variation measured from $20$ realizations of
the GMRT simulations. The dark shaded region shows the cosmic variance
which has been calculated by setting the system noise $\sigma_n=0$ in the simulation, 
 and the error bars show $1\sigma$ error bars 
predicted using eq.~(\ref{eq:be6}). The errors are dominated by the cosmic variance 
at $\ell \le 2,500$ where the dark and faint shaded regions coincide. 
We see that the Bare Estimator correctly 
recovers the input model, and the predicted error bars are consisted with 
the errors measured from the simulations.
}
\label{fig:ps1st}
\end{center}
\end{figure}

The system noise only appears in the diagonal elements of the visibility correlation matrix ${\bf V}_2$,
whereas the sky signal contributes to both the diagonal and the off-diagonal elements. Further, the diagonal
elements of the weight matrix ${\bf w}$ are all zero. Consequently the trace $Tr({\bf w}{\bf V}_2)$
in eq.~(\ref{eq:be3})  does not
pick up any contribution from the diagonal elements of  ${\bf V}_2$, and the expectation value of the 
estimator is not affected by the system noise. The variance $\sigma^2_{E_B}(a)$ however has contributions
from both diagonal and off-diagonal elements of ${\bf V}_2$. The diagonal elements  are dominated by the 
system noise, whereas the off-diagonal elements contribute to the cosmic variance. 

The weights $w_{ij}$ should, in principle, be chosen so as to maximize the signal 
to noise ratio ${\rm SNR}=\langle \hat E_B(a) \rangle/\sigma_{E_B}(a)$.  The optimal 
weights depend on the baseline distribution and $V_0 C_{\ell}/\sigma_n^2\,$, 
the relative amplitude of   the  signal to the noise  in the individual visibilities.  
Here we have made the simplifying assumption that all the visibility  pairs 
contribute equally to $\sigma^2_{E_B}(a)$. Each visibility pair is assigned 
the weight $w_{ij}=(1-\delta_{ij}) e^{-\mid \Delta \u_{ij} \mid^2/\sigma_0^2}$ which 
is proportional to its contribution to   $\langle \hat E_B(a) \rangle$. 

To test the  Bare Estimator we have used it to estimate $C_{\ell}$ from 
the simulated GMRT and Random data. For this analysis 
the visibilities with baselines $U$ in the range $40 \le U \le 1,000$ were 
divided in $20$ equally spaced logarithmic bins. Figure \ref{fig:ps1st}  
shows the mean and the rms. variation  of 
$\ell(\ell+1) C_{\ell}/2 \pi$ 
measured from the  $20$ independent realizations of the data. We find that 
the angular power spectrum estimated from the simulated GMRT data is in
good agreement  with the model (eq.~\ref{eq:cl150}) that was used 
to simulate the data. 
We next test  the predicted error estimate  $\delta C_{\ell}$ 
given by eq.~(\ref{eq:be6}).  To do this 
we have evaluated  $\sigma^2_{E_B}(a)$  by explicitly carrying out the 
sum $\sum_{ijkl}$  where the indices each runs over all the baselines
in bin $a$. For ${\bf V}_2$ (eq.~\ref{eq:vcorr}) we have used the mean 
$C_{\ell}$ estimated from the $20$ realizations and the value of 
$\sigma_n$ that was used  for the system noise in the simulation. We
find that $\delta C_{\ell}$ predicted by the analytic error estimate 
(eq.~\ref{eq:be6}) is in reasonably good agreement with the rms.
obtained from the $20$ independent realizations of the data.  
The results for the Random data are very similar to those for GMRT, 
and we have not shown these separately here. 

In conclusion of this section we find that the Bare Estimator 
(eq.~\ref{eq:be1}) is able to successfully extract the angular power 
spectrum directly from the measured visibilities. We further show that 
(eq.~\ref{eq:be6}) provides  a reasonably good estimate of the statistical 
errors for the measured angular power spectrum. The errors depend 
on the choice of the weights $w_{ij}$, the baseline distribution, the 
magnitude of the signal  and the system noise.   In  Figure \ref{fig:ps1st} 
we  see that the error decreases
with increasing $\ell$ until $\ell \sim 2,500$ beyond which the error 
increases again. We find that this feature does not change significantly
between the GMRT and the Random simulations. Based on this we 
conclude that this behaviour of the error is largely determined 
by the relative contributions from the signal whose magnitude 
falls with $\ell$ and the system noise which has been  assumed to be 
constant across all baselines. 
The errors at $\ell \le 2,500$ are cosmic variance dominated, 
whereas the errors are dominated by the system noise at larger 
$\ell$.  

\section{The Tapered  Gridded Estimator}
\label{sec:grid}
The telescope primary beam 
is usually not very well quantified at large angles 
where we have the frequency dependent pattern of nulls and sidelobes
(Figure~\ref{fig:gauss}). Point sources
located near the nulls and the sidelobes are a problem for estimating 
the angular power spectrum of the diffuse background radiation. 
Further, point sources located far away from the pointing center, 
particularly those located near the nulls, introduce ripples along the 
frequency 
direction in the multi-frequency angular power spectrum.
This poses a severe problem for separating the foregrounds 
from the cosmological 21-cm signal. As pointed out in \citet{ghosh2}, 
it is possible to avoid these problems  by tapering the sky response
 through 
a frequency independent window function ${\cal W}(\theta)$.  In this work we
choose a Gaussian ${\cal W}(\theta)=e^{-\theta^2/\theta^2_w}$  
such that $\theta_w = f \theta_0$
with $f \le 1$ so that the window function  cuts off the sky response 
well  before the first null. This tapering is achieved by convolving 
the measured visibilities 
\begin{equation}
\V_c(\u)=\tilde{w}(\u)\otimes \V(\u)
\label{eq:ge1}
\end{equation}
where $\tilde{w}(\u)=\pi\theta_w^2e^{-\pi^2U^2\theta_w^2}$ is the Fourier 
transform of ${\cal W}(\theta)$. 
The convolved visibilities $\V_c(\u)$ are the Fourier transform 
of the product ${\cal W}(\theta)\,  {\cal A}(\theta) \, \delta I(\th)$
whose  sky response  can be well controlled through the window 
function ${\cal W}(\theta)$. 

Current  radio interferometers are expected to produce considerably 
large volumes of visibility data in observations spanning many 
frequency channels and large observing times.  Given the potentially 
large computational requirement,  it is  useful to compress the 
visibility data by gridding it.  We choose a rectangular grid
in the $uv$ plane and consider the convolved visibilities  
\begin{equation}
\V_{cg} = \sum_{i}\tilde{w}(\u_g-\u_i) \, \V_i
\label{eq:ge2}
\end{equation}
where $\u_g$ refers to the different grid points and $\V_i$ 
refers to the measured visibilities. We now focus our attention 
on $\S_{cg}=\sum_{i}\tilde{w}(\u_g-\u_i) \, \S_i$ which is the sky
signal contribution to $\V_{cg}$. This can be written as 
\begin{equation}
\S_{cg}=\int d^2 U \, \tilde{w}(\u_g-\u) B(\u) \S(\u)
\label{eq:ge3}
\end{equation}
where $B(\u)=\sum_i \delta^2_D(\u-\u_i)$ is the  baseline sampling function
of the measured visibilities and $\delta^2_D(\u)$ is the 2D Dirac delta
function. The integral in eq.~(\ref{eq:ge3}) is dominated by the contribution
from within a disk of radius $\sim (\pi \theta_w)^{-1}$ centered
around $\u_g$. Assuming that the sampling function  $B(\u)$ is nearly uniform 
within this disk  we can  replace $B(\u)$ in eq.~(\ref{eq:ge3}) by its 
average value 
\begin{equation}
\bar{B}(\u_g)=\left[\frac{\int d^2 U \, \tilde{w}(\u_g-\u) B(\u)}
{\int d^2 U \, \tilde{w}(\u_g-\u)} \right]
\label{eq:ge4a}
\end{equation}
 evaluated at the grid point $\u_g$.  We  then have the approximate equation 
\begin{equation}
\S_{cg}= \bar{B}(\u_g) \int d^2 U \, \tilde{w}(\u_g-\u)  \S(\u) \,.
\label{eq:ge4}
\end{equation}
Considering eq.~(\ref{eq:ge4a}) for $\bar{B}(\u_g)$, the denominator has value ${\cal W}(0)=1$  whereby
$\bar{B}(\u_g)=\sum_i  \tilde{w}(\u_g-\u_i)$ and we have 
\begin{equation}
\S_{cg}=\left[\sum_i  \tilde{w}(\u_g-\u_i)  \right] \int d^2 U \, \tilde{w}(\u_g-\u)  \S(\u) \,. 
\label{eq:ge5}
\end{equation}
We note that eq.~(\ref{eq:ge5})  holds only if we have an uniform  and sufficiently dense 
baseline distribution in the vicinity of the
grid point $\u_g$. This breaks down if we have a patchy and sparse baseline distribution, and it is 
then necessary to use
 \begin{equation}
\S_{cg}=\sum_i  \tilde{w}(\u_g-\u_i)   \S(\u_i) \,. 
\label{eq:ge5b}
\end{equation}
 In such a situation  it is necessary to take the exact patchy $uv$ distribution  into account, and  it is 
difficult to make generic analytic predictions. Here  we have assumed an uniform 
baseline distribution,  and we have used eq.~(\ref{eq:ge5})  extensively in the subsequent calculations.

The integral in eq.~(\ref{eq:ge5})  is the Fourier  transform of the product 
${\cal W}(\theta)\,  {\cal A}(\theta) \, \delta I(\th) \equiv 
 {\cal A_W}(\theta) \, \delta I(\th)$. We may think of $ {\cal A_W}(\theta)$
as a modified primary beam pattern which has a new $\theta_{\rm FWHM}$
which is a factor $f/\sqrt{1+f^2}$ smaller than  $\theta_{\rm FWHM}$
given in Table~\ref{tab:1} and whose sidelobes are strongly suppressed. 
We can approximate the modified primary beam pattern  as a Gaussian 
$ {\cal A_W}(\theta)=e^{-\theta^2/\theta_1^2}$ with $\theta_1=f (1+f^2)^{-1/2} \theta_0$.
Using this, we can generalize eq.~(\ref{eq:vcorr}) to calculate  the correlation 
of the gridded visibilities 
$V_{c2g g^{'}} = \langle \V_{c g}  \V^{*}_{c g^{'}} \rangle$. 
The crucial point is that we have to replace $V_0$ and $\sigma_0$ in  eq.~(\ref{eq:vcorr}) 
with $V_1= \frac{\pi \theta_1^2}{2} \left( \frac{\partial B}{\partial T}\right)^{2}$ and
$\sigma_1 =  f^{-1} \sqrt{1+f^2} \sigma_0$ in order to account for the modified primary beam
pattern ${\cal A_W}(\theta)$.  We then have 

\begin{equation}
V_{c2g g^{'}} = K_{1g} K^{*}_{1 g^{'}} V_1 e^{-\mid \Delta \u_{g g^{'}} \mid^2/\sigma_1^2} C_{\ell_g} +
2 \sigma_n^2 K_{2 g g^{'}} 
\label{eq:ge7}
\end{equation}

where $\ell_g=2 \pi U_g$, $K_{1g}=\sum_i  \tilde{w}(\u_g-\u_i)$, 
$K_{2g g^{'} }=\sum_i  \tilde{w}(\u_g-\u_i) \tilde{w}^{*}(\u_{g^{'}}-\u_i)$
and  $\Delta \u_{g g^{'}}= \u_{g}-\u_{g^{'}}$.

We now define the estimator ${\hat E}_g$ for the angular power spectrum at a single grid point 
$g$ as 
\begin{equation}
{\hat E}_g=\frac{(\V_{cg} \V^{*}_{cg} - \sum_i \mid  \tilde{w}(\u_g-\u_i) \mid^2 \,  
\mid \V_i \mid^2)}{(\mid K_{1g} \mid^2 V_1 - K_{2gg} V_0)} \,.
\label{eq:ge8}
\end{equation}
Using eq.~(\ref{eq:ge7}) and eq.~(\ref{eq:vcorr})   respectively to evaluate the  expectation values 
\begin{equation}
\langle \V_{cg} \V^{*}_{cg}  \rangle =  \mid K_{1g} \mid^2  V_1 C_{\ell_g} + 2 \sigma_n^2 K_{2 g g} 
\end{equation}
and 
\begin{equation}
 \sum_i \mid  \tilde{w}(\u_g-\u_i) \mid^2 \,  \langle \mid \V_i \mid^2 \rangle 
=  V_0 \sum_i \mid  \tilde{w}(\u_g-\u_i)\mid^2  C_{\ell_i} + 2 \sigma_n^2  K_{2gg} 
\end{equation}
we see that the system noise contributions to these two terms are exactly equal and it exactly 
cancels out in  $\langle {\hat E}_g \rangle$. Further, assuming that 
$ \sum_i \mid  \tilde{w}(\u_g-\u_i) \mid^2 C_{\ell_i}   \approx C_{\ell_g} K_{2gg}$ we have 
\begin{equation}
\langle {\hat E}_g \rangle = C_{\ell_g} \,.
\end{equation}
We see that ${\hat E}_g$ defined in eq.~(\ref{eq:ge8}) gives an unbiased  estimate of the angular
power spectrum $C_{\ell}$ avoiding the positive noise bias caused by the  system noise. 

The terms $K_{1g}$ and $K_{2gg}$ in eq.~(\ref{eq:ge8}) are both proportional to $N_g$ the number of visibilities 
that contribute to the grid point $g$.  For large  $N_g$   it is reasonable to assume that   $\mid K_{1g} \mid^2 \gg K_{2gg}$
and  we thereby  simplify   eq.~(\ref{eq:ge8}) to obtain 
\begin{equation}
{\hat E}_g=\frac{(\V_{cg} \V^{*}_{cg} - \sum_i \mid  \tilde{w}(\u_g-\u_i) \mid^2 \,  
\mid \V_i \mid^2)}{\mid K_{1g} \mid^2V_1}
\label{eq:ge9}
\end{equation}
for the estimator. 

 We  use this to define the binned Tapered Gridded Estimator 
\begin{equation}
{\hat E}_G(a) = \frac{\sum_g w_g  {\hat E}_g}{\sum_g w_g } \,.
\end{equation}
where $w_g$ refers to the weight assigned to the contribution from any particular 
grid point. This has an expectation value 
\begin{equation}
\langle {\hat E}_G(a) \rangle = \frac{ \sum_g w_g C_{\ell_g}}{ \sum_g w_g}
\label{eq:ge10}
\end{equation}
which can be written as 
\begin{equation}
\langle \hat E_G(a) \rangle = \bar{C}_{\bar{\ell}_a} 
\label{eq:ge11a}
\end{equation}
where $ \bar{C}_{\bar{\ell}_a}$ is the average  angular power spectrum  at 
 \begin{equation}
\bar{\ell}_a =
\frac{ \sum_g w_g \ell_g}{ \sum_g w_g}
\label{eq:ge11}
\end{equation}
which is the   effective angular multipole  for bin $a$. 

We next calculate the variance of $\hat E_G(a)$ defined as 
\begin{equation}
[\delta C_{\ell_a}]^2 \equiv \sigma^2_{E_G}(a)=\langle \hat E^2_G(a) 
\rangle - \langle \hat E_G(a) \rangle^2\,.
\label{eq:ge12}
\end{equation} 
 Explicitly using eq.~(\ref{eq:ge9}) yields a rather unwieldy expression 
which is not very useful for making analytic predictions  for the variance. 
The first term   in the  numerator  of eq.~(\ref{eq:ge9})   which is of  order $N_g^2$ 
 makes a much larger  contribution to the variance than the second term  
$\sum_i \mid  \tilde{w}(\u_g-\u_i) \mid^2 \,   \mid \V_i \mid^2$ which is of order $N_g$.
In our analysis  we make the simplifying assumption  that we can drop the 
second term which yields 
\begin{equation}  
\sigma^2_{E_G}(a) = \frac{\sum_{g g^{'}} w_g w_g^{'}  \mid K_{1g}^{-1}   K_{1g^{'}}^{*-1} 
 V_{c2g g^{'}} \mid^2}{V_1^2[\sum_{g } w_g]^2} \,.
\label{eq:ge13}
\end{equation}
We further approximate $K_{2g g^{'}}= e^{-\mid \Delta \u_{g g^{'}} \mid^2/\sigma_1^2} K_{2gg}$
which allows us to write the variance as 
\begin{equation}  
\sigma^2_{E_G}(a) = \frac{\sum_{g g^{'}} w_g w_g^{'} e^{-2 \mid \Delta \u_{g g^{'}} \mid^2/\sigma_1^2}
\mid  C_{\ell_g} + \frac{2 K_2g g^{'} \sigma_n^2}{K_{1g} K^{*}_{1 g^{'}} V_1}  \mid^2}
{[\sum_{g } w_g]^2}
\label{eq:ge14}
\end{equation}
using eq.~(\ref{eq:ge7}).

\begin{figure}
\begin{center}
\psfrag{cl}[b][b][1.5][0]{$\ell (\ell+1) C_{\ell}/2 \pi \, [mK^2]$}
\psfrag{U}[c][c][1.5][0]{$\ell$}
\psfrag{Ghosh}[cr][tr][1.2][0]{$C^M_{\ell}$}
\psfrag{gmrt}[r][r][1.2][0]{GMRT}
\psfrag{rndm}[r][r][1.2][0]{Random}
\includegraphics[width=80mm,angle=0]{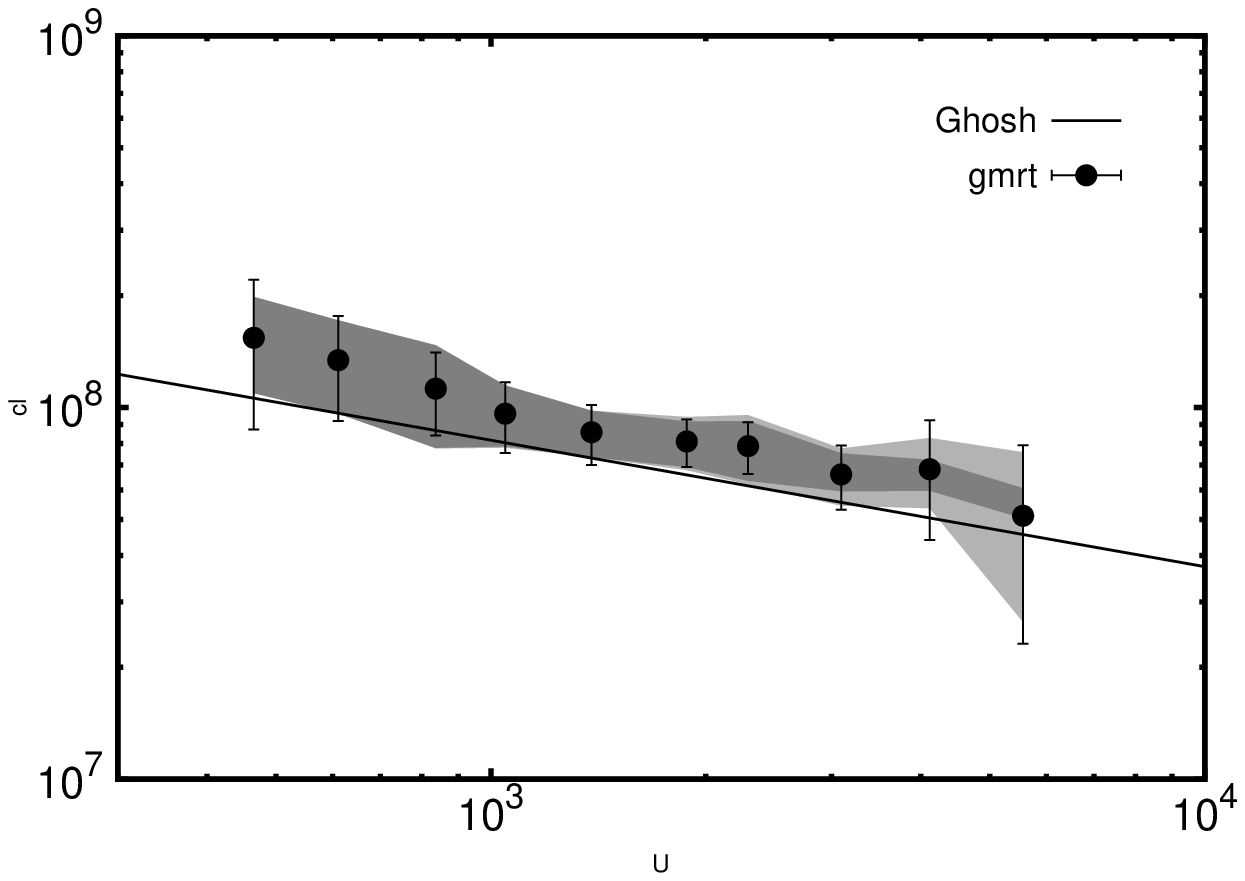}
\includegraphics[width=80mm,angle=0]{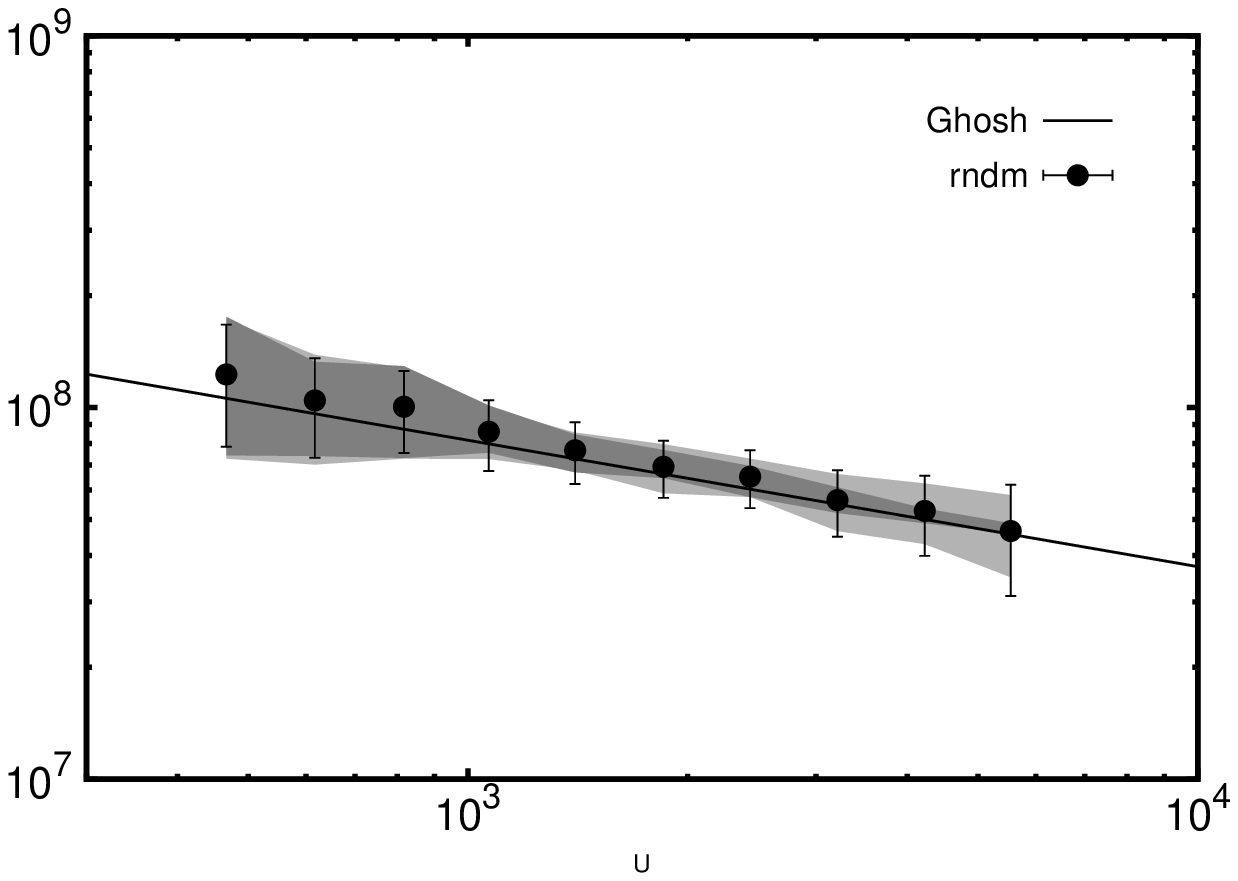}
\caption{Same as Figure \ref{fig:ps1st}, but for the Tapered Gridded Estimator.}
\label{fig:ps2nd}
\end{center}
\end{figure}

We have applied the Tapered Gridded Estimator to the simulated GMRT and Random data.
The $20$ realizations were used to calculate the mean and the variance of the estimated $C_{\ell}$.
We have considered the values $f=1.0,0.8, 0.65$ and $0.4$ for the tapering window, and have 
also tried two different weight schemes $w_g=1$ and $w_g=K^2_{1g}$ respectively. The former assigns
 equal weight to every grid point that has same data, this is expected to minimize the cosmic variance.
 The latter scheme assigns a larger weight to grid points which have a denser visibility sampling
 relative to the grid points with sparser sampling. This is expected to minimize the system noise
 contribution.  The grid spacing $\Delta U$ in the $uv$ plane
is chosen based on two considerations. A very small value of $\Delta U$ results in a very large number
of grid points which do not contain independent signal contributions. This also unnecessarily increases 
the computation time. In contrast, a large value of $\Delta U$ implies that the signal in many visibilities 
is very poorly represented in the gridded data, resulting in  a loss of signal.   We have chosen a  grid spacing
 $\Delta u=\sqrt{\ln2}/(2\pi\theta_w)$ which corresponds to one fourth of the FWHM of $\tilde{w}(\u)$ as an 
optimum value. For any fixed grid position $\u_g$, we have restricted the contribution to baselines $\u_i$ 
 within $\mid \u_g - \u_i \mid \le 6 \Delta U$. The weight function  $\tilde{w}( \u_g - \u_i)$ falls considerably
and we do not expect a significant contribution from the visibilities beyond this baseline  separation. 
The tapering also modifies the smallest baseline where the approximation of eq.~(\ref{eq:v6a}) is valid, 
and the grid points $\u_g$ in the range $U_{min}=\sqrt{1+f^2} f^{-1} 40$ to $1,000$ were binned into 
$10$ equally spaced logarithmic bins for this analysis.

Figure \ref{fig:ps2nd} shows the results for $f=0.8$ and $w_g=\mid K_{1g}\mid^2$. We see that for  both 
GMRT and Random 
the estimated $C_{\ell}$ are roughly within the $1\sigma$ region of the input model angular power spectrum 
$C^{M}_{\ell}$. For GMRT, 
however, the estimated $C_{\ell}$ values all appear to be somewhat in excess of  $C^M_{\ell}$ indicating 
that we have an overestimate of  the angular power spectrum relative to $C^{M}_{\ell}$. In 
comparison, the $C_{\ell}$ values are in better agreement with $C^{M}_{\ell}$ for the Random simulation. 
For both GMRT and Random the error estimates predicted by
eq.~(\ref{eq:ge14}) are in good agreement with the rms. fluctuation
estimated from the 20 realizations.  We note  that the rms. fluctuation of $C_{\ell}$ is more for GMRT in 
comparison to Random. 

\begin{figure}
\begin{center}
\psfrag{dev}[c][c][1.5][0]{$(C_{\ell}-C^M_{\ell}) /C^M_{\ell}$}
\psfrag{U}[c][c][1.5][0]{$\ell$}
\psfrag{gmrt}[c][c][1.2][0]{GMRT}
\psfrag{rndm}[c][c][1.2][0]{Random}
\psfrag{g100}[r][r][1.][0]{100}
\psfrag{g20}[r][r][1.0][0]{20}
\psfrag{g10}[r][r][1.0][0]{$N_r=10$}
\psfrag{a100}[r][r][1.][0]{100}
\psfrag{a10}[r][r][1.][0]{$N_r=10$}
\psfrag{a20}[r][r][1.][0]{20}
\psfrag{b100}[r][r][1.][0]{100a}
\includegraphics[width=80mm,angle=0]{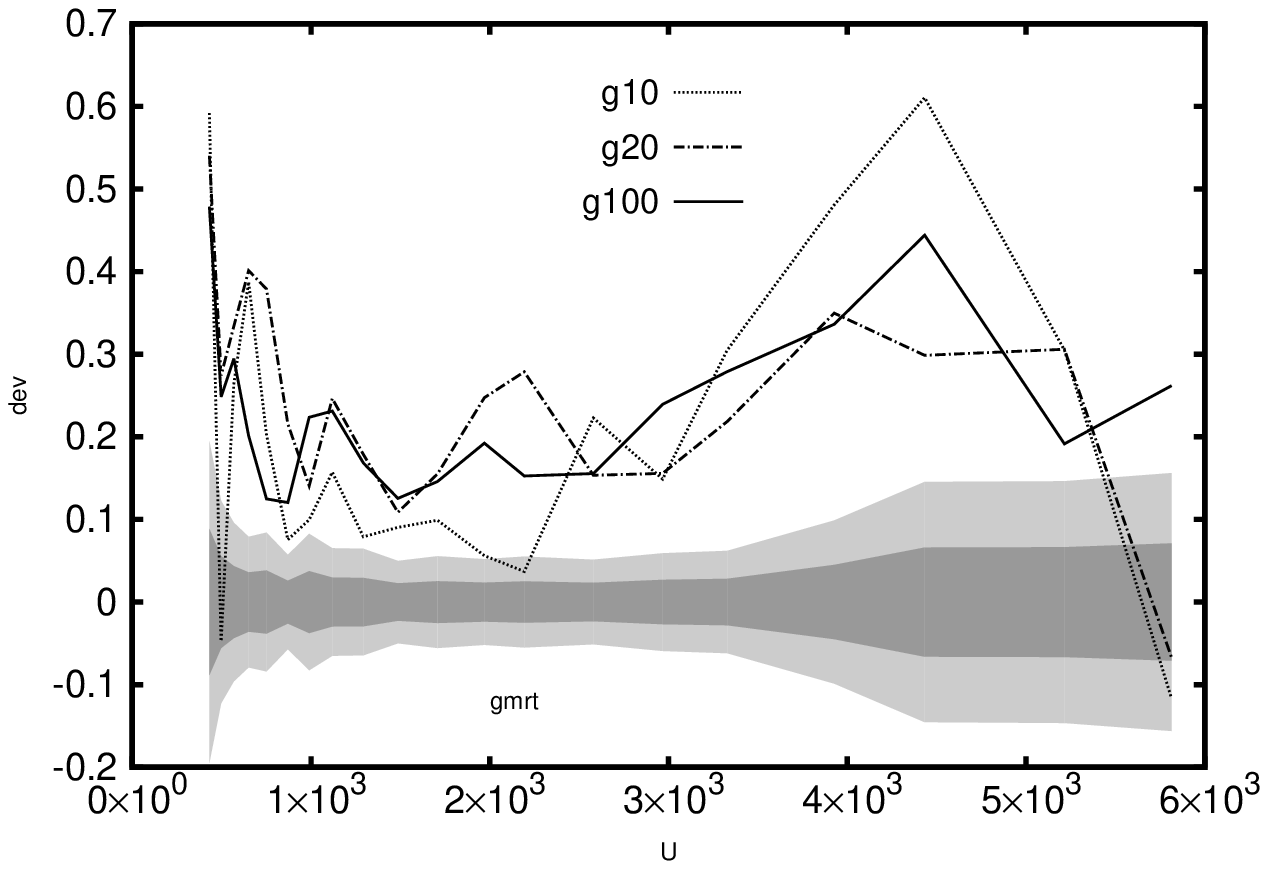}
\includegraphics[width=80mm,angle=0]{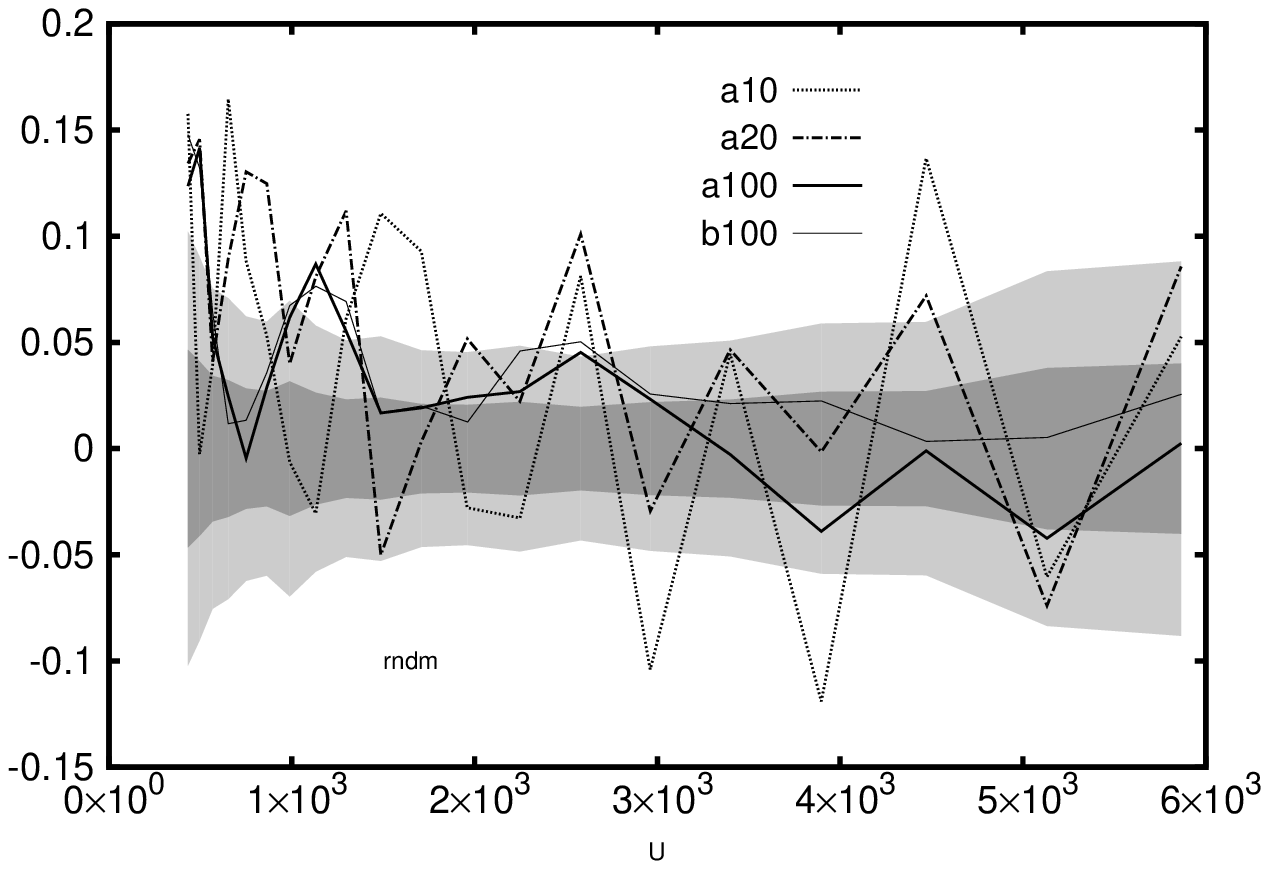}
\caption{The different curves  show  the fractional deviation  $(C_{\ell}-C^M_{\ell})/C^M_{\ell}$
for the  different numbers of realizations $(N_r)$ shown in the figure.  The  curve  100a corresponds to 
$N_r=100$ with  $869,828$ baselines,   which is  $4$ times 
the number of baselines in the other simulations. The two shaded region show  $\sigma/(\sqrt{N_r} \, C_{\ell}^M)$
for $N_r=20$ and $100$ respectively. We have used $f=0.8$ and $w_g=\mid K_{1g} \mid^2$, with $20$ equally spaced
logarithmic bins in $\ell$.}
\label{fig:dev}
\end{center}
\end{figure}

The  Tapered Gridded Estimator is  expected to give  an unbiased estimate of  $C_{\ell}$ 
provided  we have a uniform and sufficiently  dense  baseline distribution. We  test this 
using  the Random simulations which have a uniform baseline distribution.  
In such a  situation we expect the  deviation $C_{\ell} -C_{\ell}^M $ to arise purely from 
 statistical fluctuations. The deviation  is expected to have values  around  $~\sigma/\sqrt{N_r}$  
and   converge to $0$ as $N_r$, the number of  realizations,  is increased. 
For this purpose we have studied (Figure \ref{fig:dev})
how  the  fractional deviation $(C_{\ell}-C_{\ell}^M)/C_{\ell}^M$  
  varies if we increase the number of  realizations from  $N_r=10$ to  $100$. We find that it is 
more convenient to use $20$ equally spaced logarithmic bins in $\ell$ to highlight the convergence
of the fractional deviation with increasing $N_r$. Note that we have used $10$ bins (as mentioned earlier)
everywhere except in (Figure \ref{fig:dev}). 
For the Random simulation (right panel), 
 we find that as the number of realizations is increased  the convergence of the  fractional 
deviation to $0$ is clearly visible  for $\ell \ge 1.2 \times 10^3$ $(U \ge 200)$.
Further, the fractional deviation is also found to be consistent with $\sigma/(\sqrt{20} \, C_{\ell}^M)$
and  $\sigma/(10 \, C_{\ell}^M)$ expected for $N_r=20$ and $100$ respectively. 
At smaller baselines, however, the behaviour is not so clear.  The approximation 
eq.~(\ref{eq:v6a}) for the  convolution and the 
approximation for the primary beam pattern each introduce around $2-5 \%$ errors 
in the estimated $C_{\ell}$ at small baselines. Further, for a uniform baseline distribution 
the bins at the smallest  $\ell$ values contain  fewer  baselines and also fewer grid points, and are 
susceptible to larger  fluctuations. The discrete $uv$ sampling due to the finite number of baselines 
is also expected to introduce some errors at all values of $\ell$. To test this effect, 
we have considered a situation where $N_r=100$ and the total number of baselines is 
increase to  $869,828$ which is  a factor of $4$ larger compared to the other simulations. 
We find that for  $\ell \ge 3 \times 10^3$ the  fractional deviation  falls from $\sim 5 \%$
to $\sim 2 \%$ when the baseline density is increased, this difference is not seen at smaller
baselines. In summary, the tests clearly show that for a uniform baseline distribution
the estimator is unbiased for  $\ell \ge 1.2 \times 10^3$.  In contrast, for the GMRT (left panel)
the fractional  deviation does not converge to $0$ as $N_r$ is increased. We see that $C_{\ell}$ 
is overestimated at all values of $\ell$. As mentioned earlier, the GMRT has a patchy $uv$ coverage  
for which eq.(\ref{eq:ge4}),  which assumes a uniform baseline distribution, breaks down.  
The overestimate is a consequence of GMRT's patchy $uv$ coverage, and is not inherent  to the  Tapered 
Gridded Estimator. The rms. fluctuations also are larger for GMRT in comparison to the Random simulations
(Figure \ref{fig:ps2nd}).  This too  is a consequence of GMRT's patchy $uv$ coverage.

\begin{figure}
\begin{center}
\psfrag{cl}[b][b][1.5][0]{$ (C_{\ell}-C^M_{\ell}) /C^M_{\ell}$}
\psfrag{U}[c][c][1.5][0]{$\ell$}
\psfrag{f1wt2}[c][c][1.2][0]{f=1}
\psfrag{f2wt2}[r][r][1.2][0]{f=0.8}
\psfrag{f3wt2}[r][r][1.2][0]{f=0.65}
\psfrag{f4wt2}[r][r][1.2][0]{f=0.4}
\psfrag{gmrt}[c][c][1.2][0]{GMRT}
\psfrag{rndm}[c][c][1.2][0]{Random}
\includegraphics[width=80mm,angle=0]{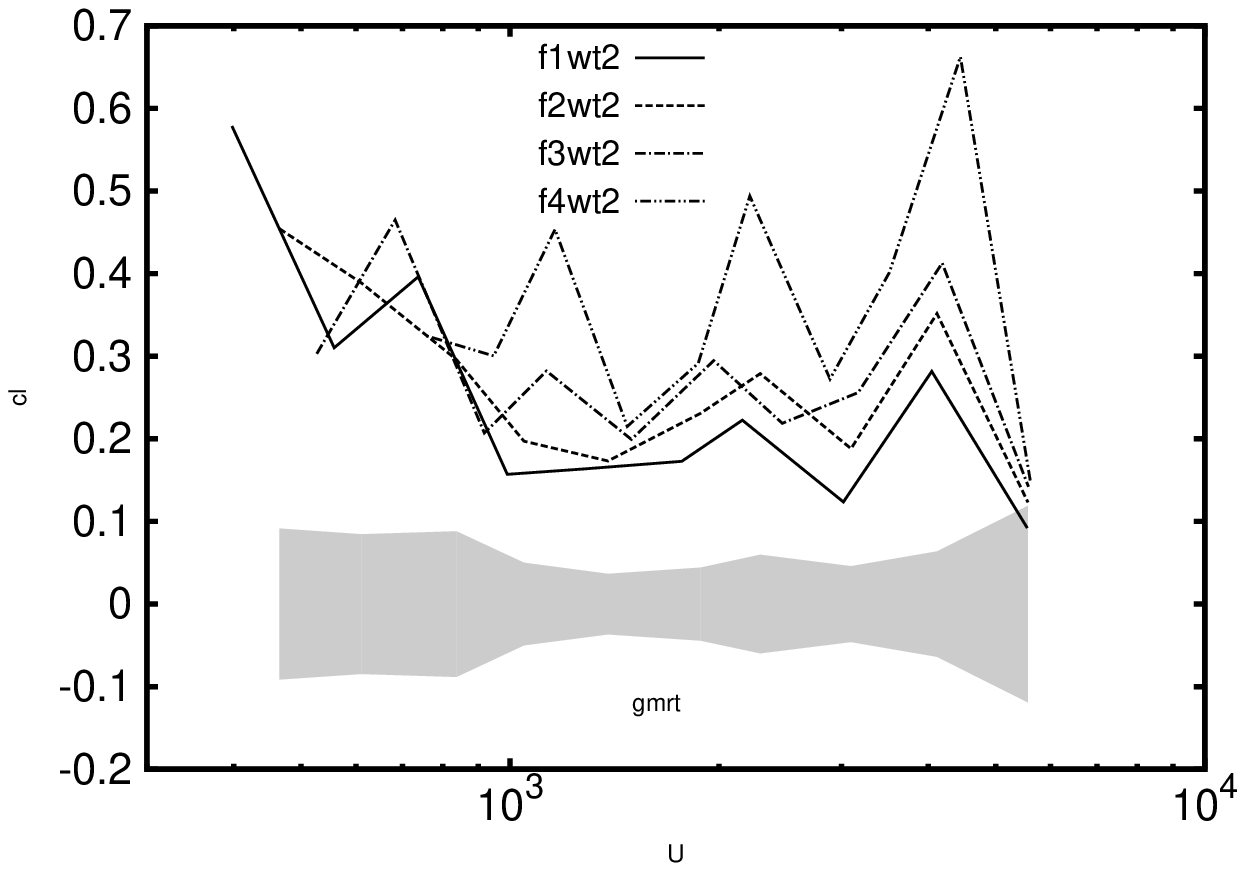}
\includegraphics[width=80mm,angle=0]{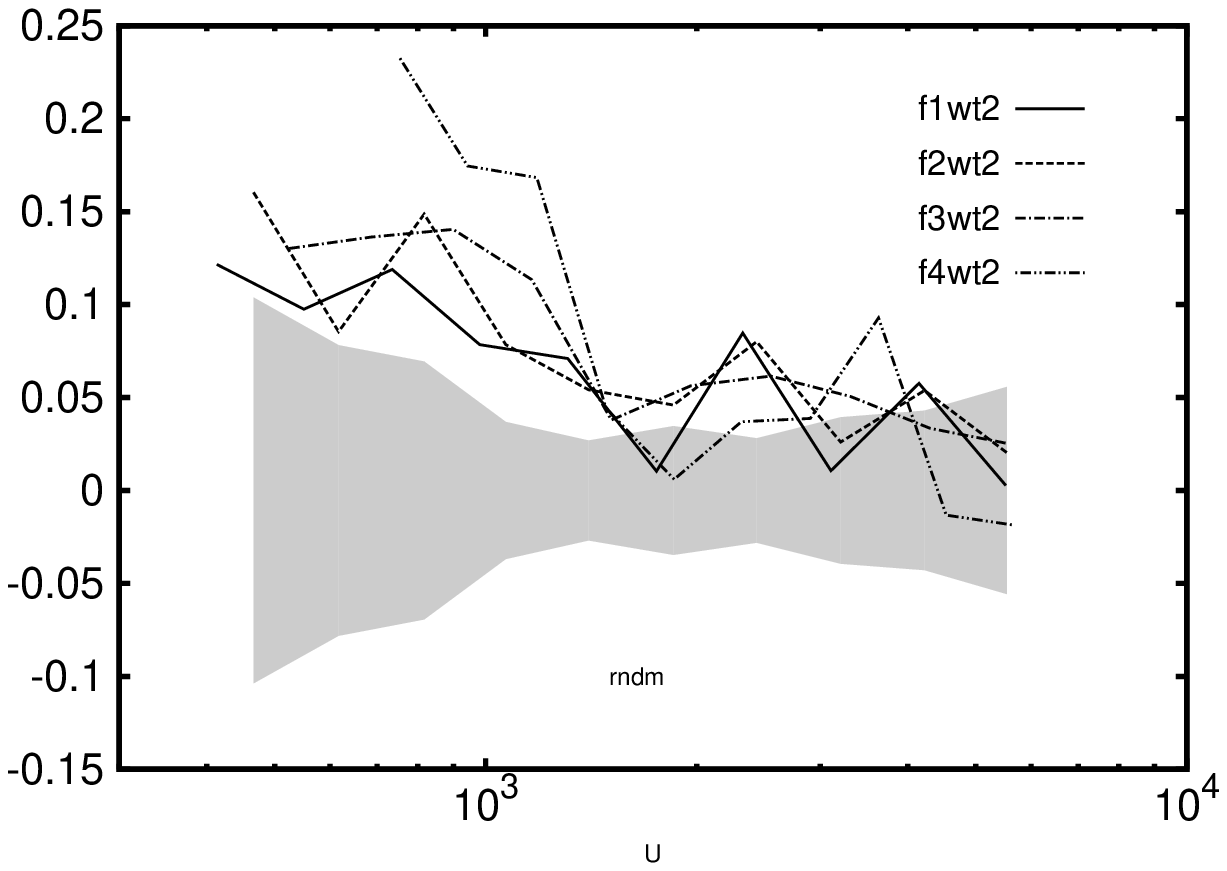}
\caption{This shows how the fractional deviation  varies with $f$ for $w_g=\mid K_{1g} \mid^2$. 
The results  are averaged over 20 realizations  of the sky signal. For comparison,   
the shaded region shows  $\sigma/(\sqrt{20} \, C_{\ell}^M)$ which is   the expected statistical fluctuation 
for $f=0.8$. }
\label{fig:dev2ndwt2}
\end{center}
\end{figure}

\begin{figure}
\begin{center}
\psfrag{cl}[b][b][1.5][0]{$\sigma/C^M_{\ell}$}
\psfrag{U}[c][c][1.5][0]{$\ell$}
\psfrag{f1wt2}[c][c][1.2][0]{f=1}
\psfrag{f2wt2}[r][r][1.2][0]{f=0.8}
\psfrag{f3wt2}[r][r][1.2][0]{f=0.65}
\psfrag{f4wt2}[r][r][1.2][0]{f=0.4}
\psfrag{gmrt}[c][c][1.2][0]{GMRT}
\psfrag{rndm}[c][c][1.2][0]{Random}
\includegraphics[width=80mm,angle=0]{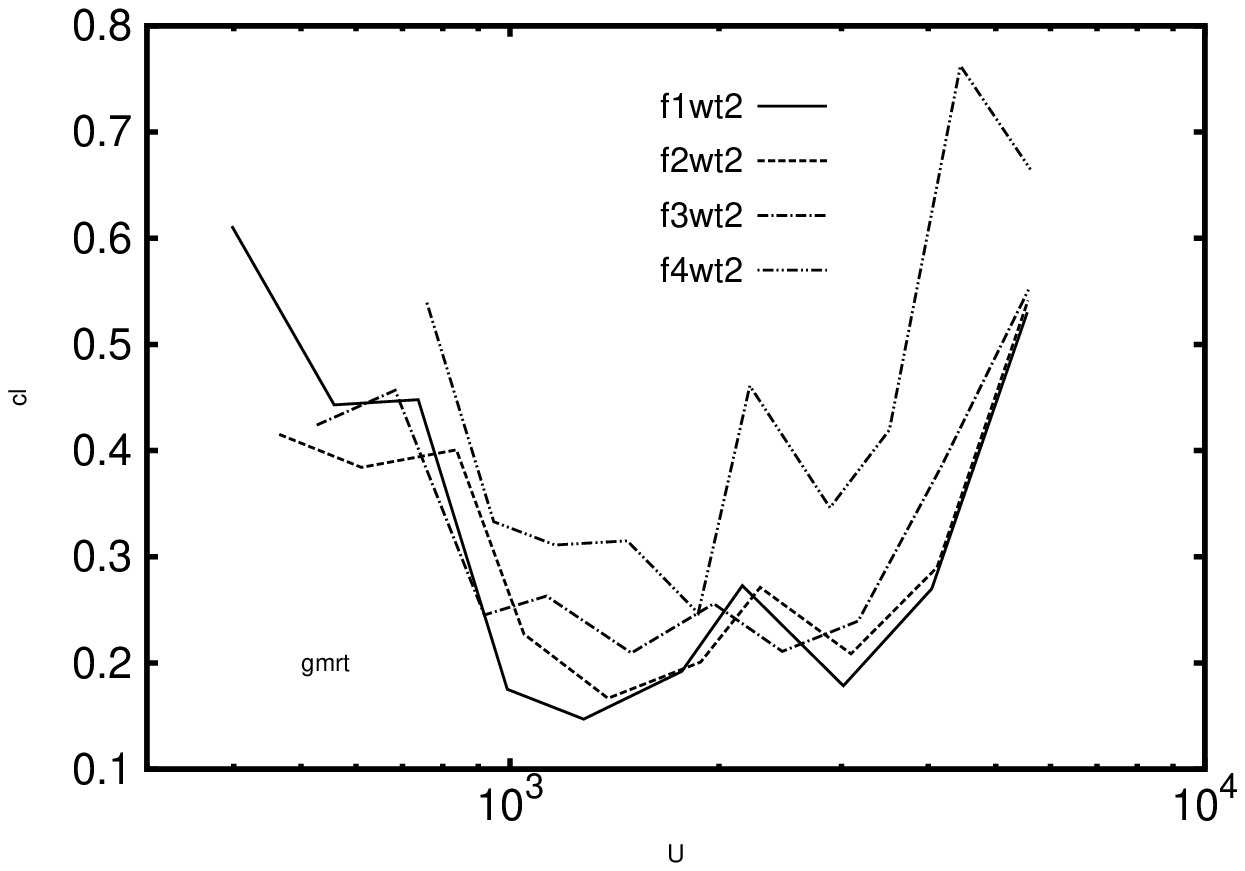}
\includegraphics[width=80mm,angle=0]{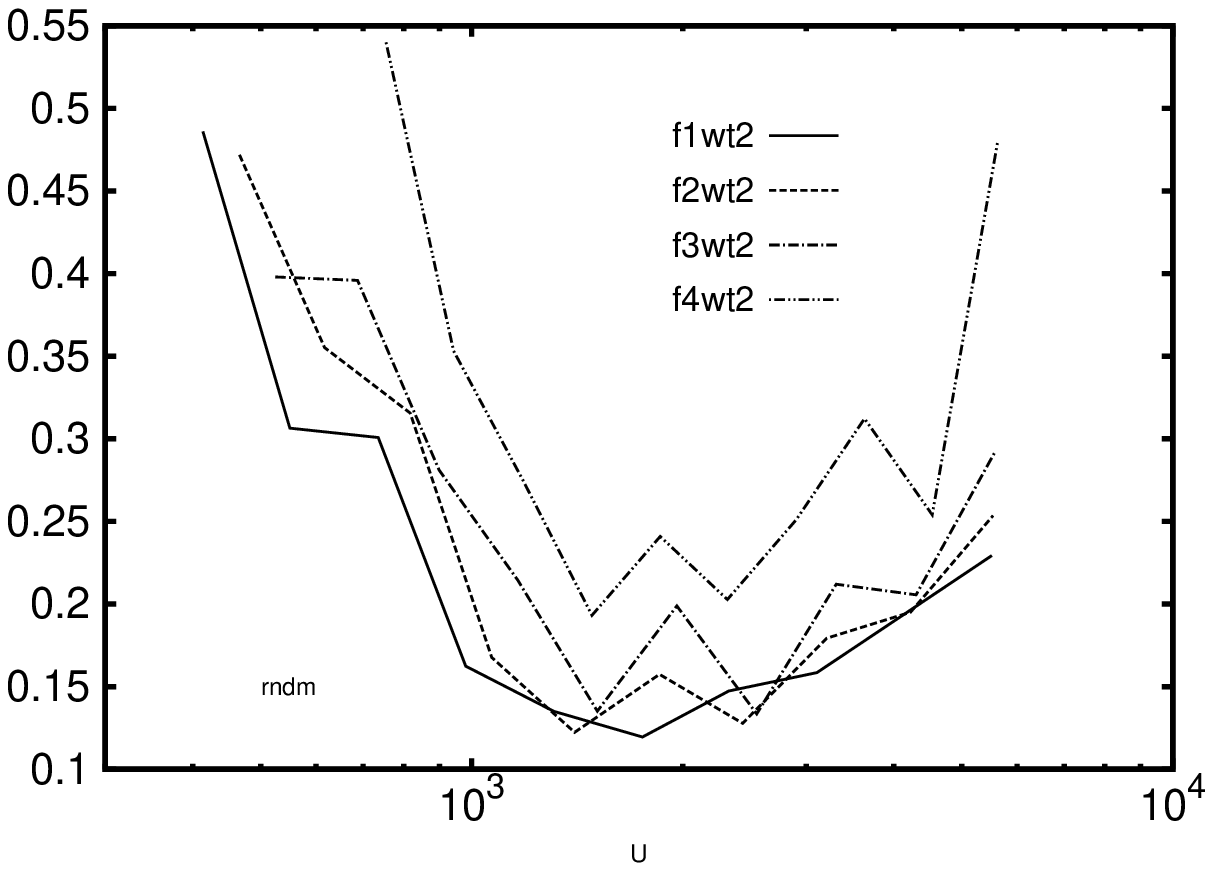}
\caption{This figure shows the relative error ($\sigma/C^M_{\ell}$) estimated from 20 realization of the simulation. Here, we have used $w_g=\mid K_{1g} \mid^2$ and the  different $f$ values shown in this figure.}
\label{fig:sig2ndwt2}
\end{center}
\end{figure}

We now study how the estimator behaves for different values of
$f$. Figure \ref{fig:dev2ndwt2} and Figure \ref{fig:sig2ndwt2}
respectively show the relative deviation $(C_{\ell}-C^M_{\ell})
/C^M_{\ell}$ and the relative error $\sigma/C^M_{\ell}$ for different
values of $f$ with $w_g=\mid K_{1g} \mid^2$. Here, $C_{\ell}$ and
$\sigma$ refer to the mean and rms. estimated from the 20
realizations. We find that the deviations are roughly within the
$1\sigma$ errors for all the cases that we have considered.  For
GMRT, the deviation increases with decreasing $f$. This effect is only
visible at low $\ell$ for Random. The error $\sigma$, increases with
$f$ for both GMRT and Random. In all cases, the error is found to
decrease until $\ell\sim2000$ and then increase subsequently. As
mentioned earlier for the Bare Estimator, we interpret this as a
transition from cosmic variance to system noise dominated errors as
$\ell$ is increased. The sky coverage of the modified primary beam
${\cal A_W}(\theta)$ falls with a decrease in $f$. This explains the
bahaviour of the cosmic variance contribution which increases as $f$
is reduced. We further see that the system noise contribution also
increases as $f$ is reduced. This can be attributed to the term
$V_1=\frac{\pi\theta_1^2}{2}$ which appears in
eq.~(\ref{eq:ge14}). This effectively increases the system noise
contribution relative to $C_{\ell}$ as $f$ is reduced.

We have studied the relative performance of the two weight scheme
mentioned earlier. Figure {\ref{fig:sig2ndf2}} shows the relative
deviation and the relative error for both $w_g=1$ and $w_g=\mid K_{1g}
\mid^2$ for $f=0.8$. As expected, the first scheme performs better in
the cosmic variance dominated regime. The difference between the two
weight scheme, however, is not very large in this regime. The second
weight scheme performs significantly better in the system noise
dominated region. In this region the errors are nearly doubled if we
use $w_g=1$ instead of $w_g=\mid K_{1g} \mid^2.$ 

In summary, we have introduced a Gridded Estimator for the angular
power spectrum where it is possible to avoid the positive noise bias
which arises due to the contribution from the correlation of a visibility with itself.
Further, the estimator allows the possibility to taper the sky response
and thereby implement  sidelobe suppression. We have used simulated visibility data
to validate the estimator.  We find that the estimator provides an unbiased estimate
of $C_{\ell}$ for $\ell \ge 1.2 \times 10^{3}$ if we have a sufficiently 
dense, uniform baseline distribution.   We also find that eq.~(\ref{eq:ge14}) provides 
a good analytic estimate of the errors in the measured $C_{\ell}$.
The estimator is found to be sensitive to the telescope's $uv$ coverage,  
and we have somewhat of an overestimate for the GMRT which has a patchy
$uv$ coverage. This deviation, however, is roughly within the $1\sigma$ 
error bars and is not expected to be a serious issue. It is possible to carry
out  simulations with the actual observational $uv$ coverage and use these
to compensate for the overestimate. The new telescopes like LOFAR 
(discussed later) have a denser and more uniform $uv$ coverage, and we do not 
expect this issue to be of concern there. The $1\sigma$ errors, we find, increase as the 
tapering is increased. The choice of $f$, however, is decided by issues
related to point source removal not considered here. We find that the weight
scheme $w_g=\mid K_{1g} \mid^2$ performs better than $w_g=1$, and we use the former
for the subsequent analysis.

\section{A comparison of the two  estimators}
\begin{figure}
\begin{center}
\psfrag{cl}[c][c][1.5][0]{$\sigma/C^M_{\ell}$}
\psfrag{dev}[c][c][1.5][0]{$(C_{\ell}-C^M_{\ell}) /C^M_{\ell}$}
\psfrag{U}[c][c][1.5][0]{$\ell$}
\psfrag{f2}[r][l][1.2][0]{f=0.8}
\psfrag{f2wt1}[r][c][1.][0]{$w_g=1$}
\psfrag{f2wt2}[r][r][1.][0]{$w_g=\mid K_{1g}\mid^2$}
\psfrag{bare}[r][l][1.2][0]{Bare}
\psfrag{gmrt}[c][c][1.2][0]{GMRT}
\includegraphics[width=80mm,angle=0]{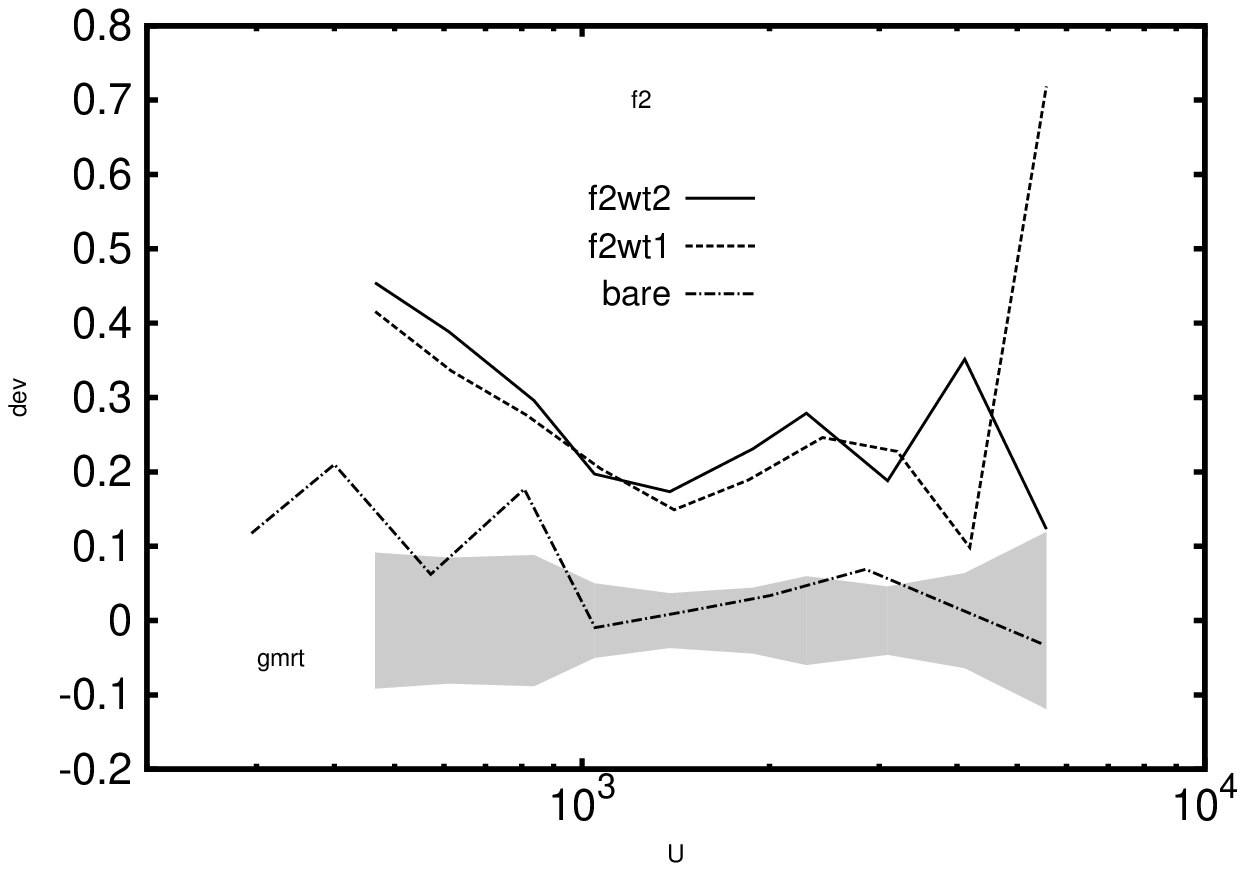}
\includegraphics[width=80mm,angle=0]{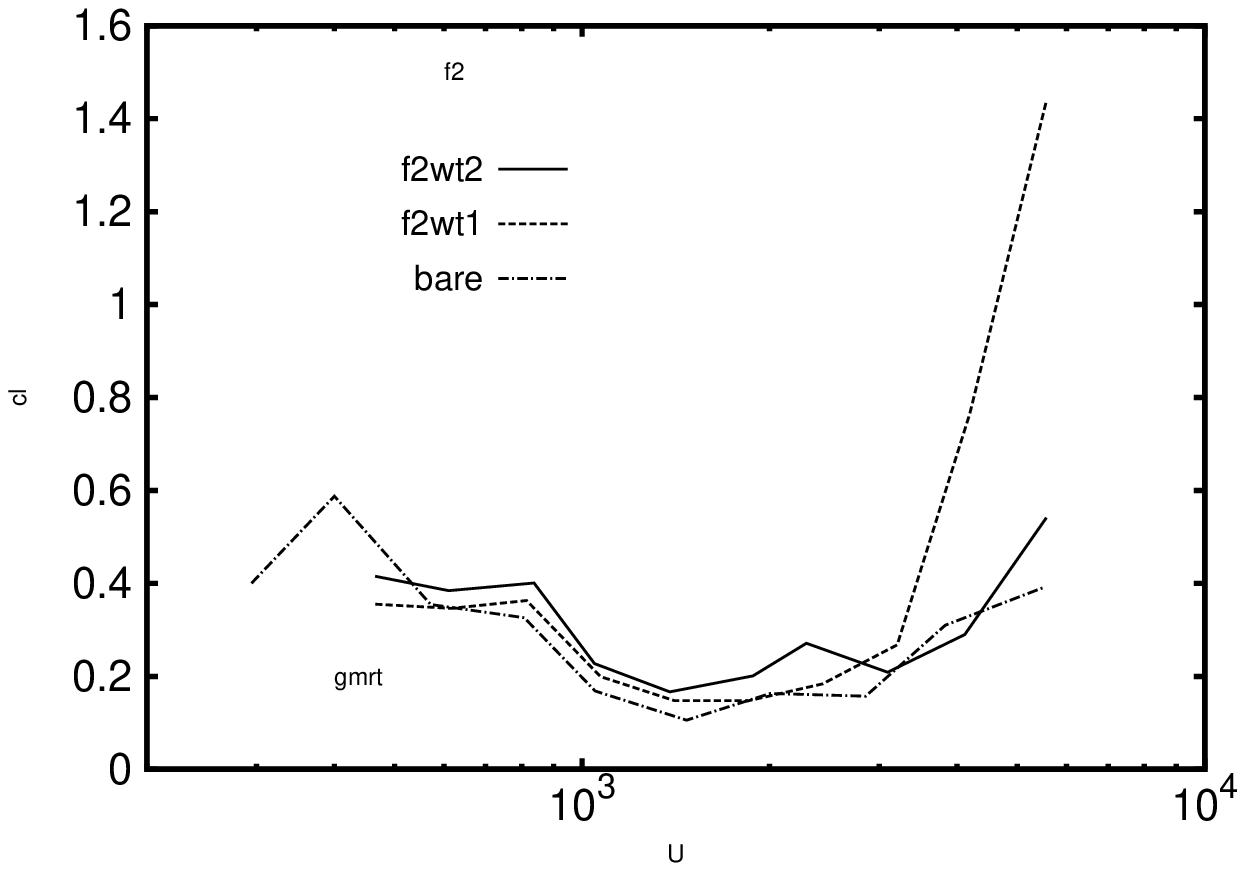}
\caption{ The left (right) panel shows the fractional deviation (error) for the two weight schemes 
 $w_g=1$ and $\mid K_{1g}\mid^2$ respectively, both with $f=0.8$.  The results for the Bare estimator have also 
been shown for comparison. The results  are based on  20 realizations  of the sky signal. 
For comparison,  the shaded region in the left panel shows  $\sigma/(\sqrt{20} \, C_{\ell}^M)$ which is
the expected statistical fluctuation  for $w_g=\mid K_{1g}\mid^2$.}
\label{fig:sig2ndf2}
\end{center}
\end{figure}

Comparing the Bare Estimator with the Tapered Gridded Estimator we see (left panel of Figure~\ref{fig:sig2ndf2}) that the former is more successful in recovering the input sky model. The statistical errors also (right panel of Figure~\ref{fig:sig2ndf2}) , we find, are somewhat smaller for the Bare Estimator. The Bare Estimator deals directly with the measured visibilities, and in a sense we expect it to outperform any other estimator which deals with gridded visibilities. What then is the motivation to consider a Gridded Estimator which is not able to recover the input model with as much accuracy as the Bare Estimator? The Bare Estimator deals directly with the visibilities and the computational time for the pairwise correlation in eq.~(\ref{eq:be1}) scales proportional to $N^2$, where $N$ is the total number of visibilities in the data. Further, the error calculation in eq.~(\ref{eq:be6}) is expected to scale as $N^4$. In contrast, the computation time is expected to scale as N for Tapered Gridded Estimator. This N dependence arises in the process of gridding the visibilities, the correlation eq.~(\ref{eq:ge9}) and the error estimate eq.~(\ref{eq:ge14}) are both independent of $N$.

\begin{figure}
\begin{center}
\psfrag{bline}[c][c][1.5][0]{No. of visibility}
\psfrag{time}[b][c][1.5][0]{Time [sec]}
\psfrag{1st}[r][r][1.5][0]{$C_{\ell}$ Bare}
\psfrag{2nd}[r][r][1.5][0]{$C_{\ell}$ Gridded}
\psfrag{1stvar}[r][r][1.5][0]{$\sigma$ Bare}
\includegraphics[width=70mm,angle=-90]{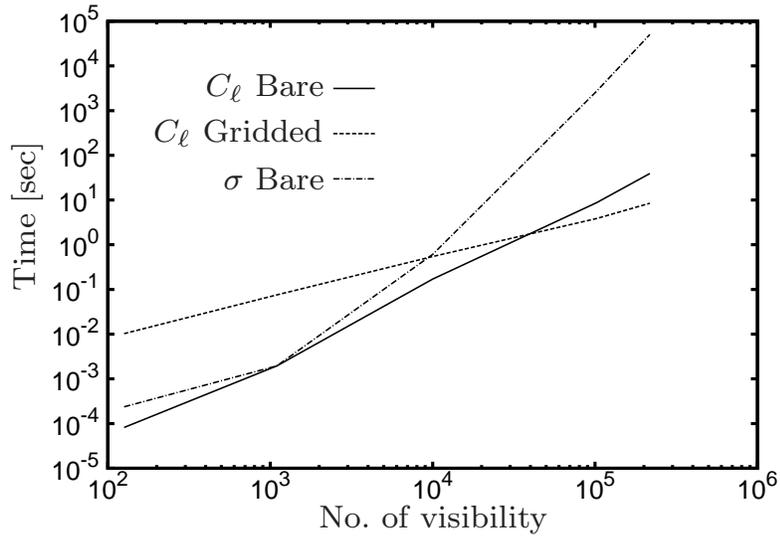}
\caption{This shows how the computation time varies with the number of visibility data  for the two different
 estimators.  The computation time for  analytically predicting  the error (eq.~\ref{eq:be6}) 
for the Bare Estimator is also shown.} 
\label{fig:time}
\end{center}
\end{figure}

Figure~\ref{fig:time} show the computation time for the two estimators as the number of visibilities varied. We see that the computation time shows the expected $N$ dependence for large values of $N(>1000)$. The Bare Estimator takes less computation time when $N$ is small ($N\le10^4$). However, the computation time for the Bare Estimator and its error estimate are larger than that for the Tapered Gridded Estimator for $N\ge10^5$. The Bare Estimator is extremely computation extensive for a large $N$ and it is preferable to use the Gridded Estimator when $N\ge10^5$. Based on this we focus on the Tapered Gridded Estimator for most of the subsequent discussion.

\section{Gain Errors}
\label{sec:gerr}
The measured visibilities have undetermined time varying gains which arise due to  the 
atmosphere, receiver system, etc. The calibration procedure attempts to determine these
gains and correct for them, but this generally leaves unknown residual gain errors in the data.
\citet{adatta,adatta10} have studied the impact of the residual gain errors on  bright source subtraction 
and place a tolerance limit  for  detecting the reionization $21 \,{\rm cm}$ signal. Recently, \citet{shaw14b} have discussed how amplifier gains affect on the detection of  the 21 cm signal from post-reionization and reionization era respectively. Here we study the 
effect of gain errors on the estimators that we have defined earlier. For this work we assume antenna
 dependent  gain errors whereby the calibrated visibilities can be written as 

\begin{equation}
\V(\u_{ab})=g_ag_b^*[\S(\u_{ab})+\N(\u_{ab})]
\end{equation}
where $a,b$ refer to the two antennas corresponding to the baseline $\u_{ab}$, and
  $g_a=(1+\alpha_a)e^{i\phi_a}$
and $g_b=(1+\alpha_b)e^{i\phi_b}$ are the respective antenna gains.
Here the $\alpha_a$s and the $\phi_a$s are 
respectively the amplitude and the phase errors of the individual 
antenna gains.
We have assumed that both $\alpha_a$ and $\phi_a$ are Gaussian random variables of zero mean and
 variance $\sigma_{\alpha}^2$ and $\sigma_{\phi}^2$ respectively. The errors are assumed to be independent 
in different antennas  and at different time instants . 

The two visibility correlation can  be
written as,
\begin{equation}
\langle \V(\u_{ab})\V^{*}(\u_{cd})\rangle=\langle g_ag_b^*g_c^*g_d\rangle 
[S_2(\u_{ab},\u_{cd})+N_2(\u_{ab},\u_{cd})]
\label{eqn:gerrcorr}
\end{equation}
where the product of the gains is to be averaged over different realizations of the gain errors $\alpha$ and $\phi$. We now have three
different possibilities which we discussed below.

{\bf Case I:} The two visibilities $\V(\u_{ab})$ and $\V(\u_{cd})$ are 
at two different time instants or they have no antenna 
in common. In this situation we have

\begin{equation}
\langle g_ag_b^*g_c^*g_d\rangle=e^{-2\sigma_{\phi}^2}.
\label{eq:gerrcs1}
\end{equation}
 
 {\bf Case II:} The two visibilities $\V(\u_{ab})$ and $\V(\u_{cd})$ are at the same time instant and have only one antenna in common. In this situation we have

\begin{equation}
\langle g_ag_b^*g_c^*g_d\rangle=(1+\sigma_{\alpha}^2)e^{-\sigma_{\phi}^2}.
\label{eq:gerrcs2}
\end{equation}

{\bf Case III:} Both  $\V(\u_{ab})$ and $\V(\u_{cd})$ referred
the same measured visibility. In this situation we have
\begin{equation}
\langle g_ag_b^*g_c^*g_d\rangle=(1+\sigma_{\alpha}^2)^2.
\label{eq:gerrcs3}
\end{equation}

The signal contribution to both the  estimators defined 
earlier is dominated by Case I, whereas the noise is dominated by
Case III. Based on this it is possible to generalize eq.~(\ref{eq:vcorr}) to 
obtain the approximate relation 
\begin{equation}
V_{2ij} = e^{-2\sigma_{\phi}^2}V_0 \, e^{-\mid \Delta \u_{ij} \mid^2/
\sigma_0^2} 
\, C_{\ell_i} + (1+\sigma_{\alpha}^2)^2\delta_{ij} 2 \sigma_n^2
\label{eq:vcorrm}
\end{equation}
which takes into account the effect of gain errors. It is also
possible to generalize eq.~(\ref{eq:ge7}) for the gridded visibilities
in a similar fashion. Using these  to calculate the effect of gain errors
 on the  estimators defined earlier,  we have  
\begin{equation}
\langle \hat E(a) \rangle = e^{-2 \sigma^2_{\phi}} \bar{C}_{\bar{\ell}_a} \,.
\label{eq:gerr1}
\end{equation}
for both the Bare and the Tapered Gridded Estimators. 
We see that  both the  estimators  are  unaffected by the error in the gain amplitude, 
 however the  phase errors cause the expectation value  of the estimator to decrease by a factor 
$e^{-2 \sigma^2_{\phi}}$.  It is quite straightforward to generalize eq.~(\ref{eq:be6}) and 
eq.~(\ref{eq:ge13}) to incorporate the effect of the gain errors in  the  variance  of the  Bare 
 and the Tapered Gridded Estimators   respectively. The main effect is that the signal contribution 
is suppressed by a factor  $e^{-2 \sigma^2_{\phi}}$ whereas the system noise contribution is 
jacked up by a factor $(1+\sigma_{\alpha}^2)^2$ (eq.~\ref{eq:vcorrm}). We consequently expect 
the ${\rm SNR}$ to remain unchanged in the cosmic variance dominated regime at low $\ell$, 
whereas  we  expect the ${\rm SNR}$ to fall  in the system noise dominated regime (large $\ell$). 
Further, we also expect the transition from the cosmic variance to the system noise dominated 
regime to shift to smaller $\ell$ values if the gain errors increase.

\begin{figure}
\begin{center}
\psfrag{cl}[b][b][1.2][0]{$\ell (\ell+1) C_{\ell}/2 \pi \, [mK^2]$}
\psfrag{U}[c][c][1.2][0]{$\ell$}
\psfrag{snr}[c][c][1.2][0]{${\rm SNR}$}
\psfrag{model}[c][l][1.][0]{$C^M_{\ell}$}
\psfrag{model60}[r][r][1.][0]{$C^M_{\ell}\times e^{-2\sigma^2_{\phi}}$}
\psfrag{a50}[c][r][1.][0]{$\sigma_{\alpha}=0.5$}   
\psfrag{p60}[r][r][1.][0]{$\sigma_{\phi}=60^{\circ}$}
\psfrag{p10}[r][r][1.][0]{$\sigma_{\phi}=10^{\circ}$}
\psfrag{gmrt}[c][c][1.2][0]{GMRT}
\psfrag{00}[r][r][1.][0]{Uncorrupted}
\psfrag{a10p10}[r][r][1.][0]{$\sigma_{\alpha}=0.1,\sigma_{\phi}=10^{\circ}$}
\psfrag{a10p60}[r][r][1.][0]{$\sigma_{\alpha}=0.1,\sigma_{\phi}=60^{\circ}$}
\psfrag{a50p10}[r][r][1.][0]{$\sigma_{\alpha}=0.5,\sigma_{\phi}=10^{\circ}$}
\psfrag{a50p60}[r][r][1.][0]{$\sigma_{\alpha}=0.5,\sigma_{\phi}=60^{\circ}$}
\includegraphics[width=80mm,angle=0]{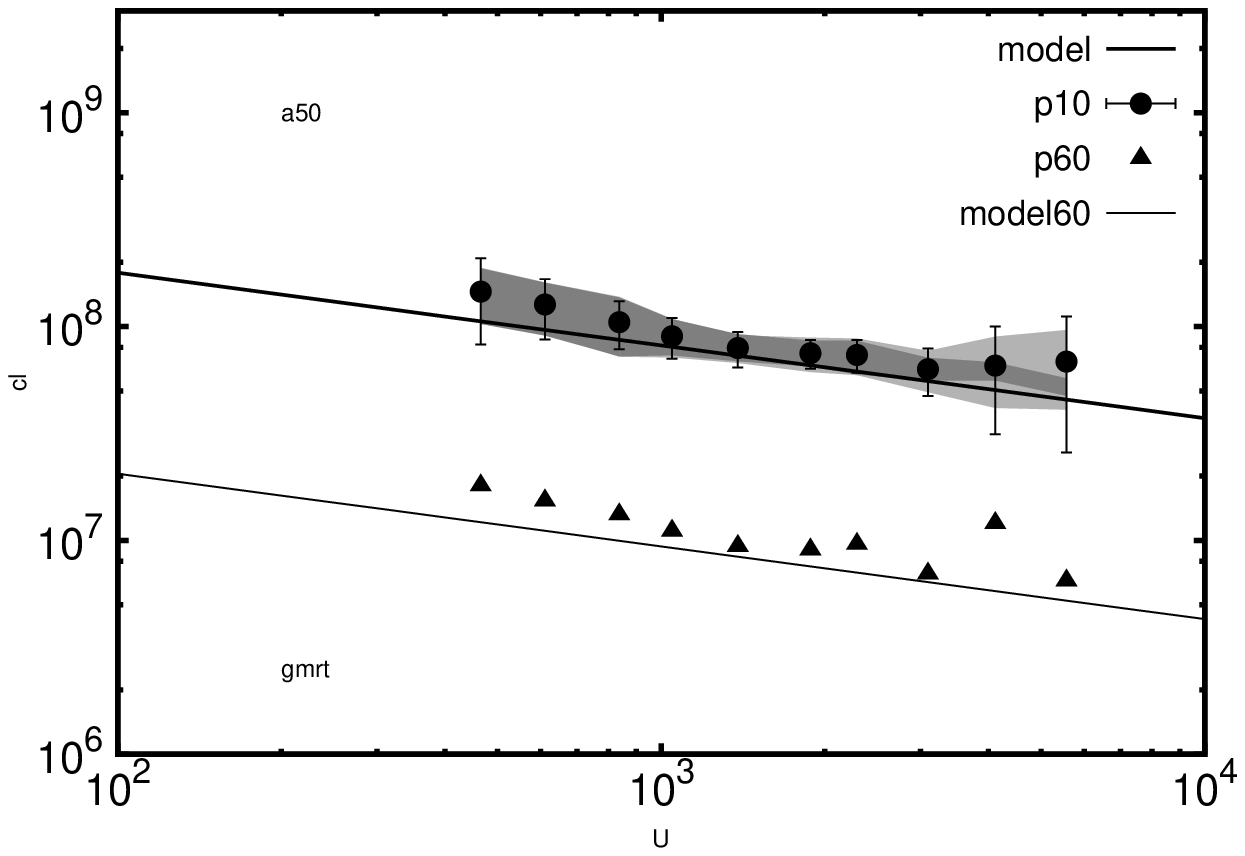}
\includegraphics[width=80mm,angle=0]{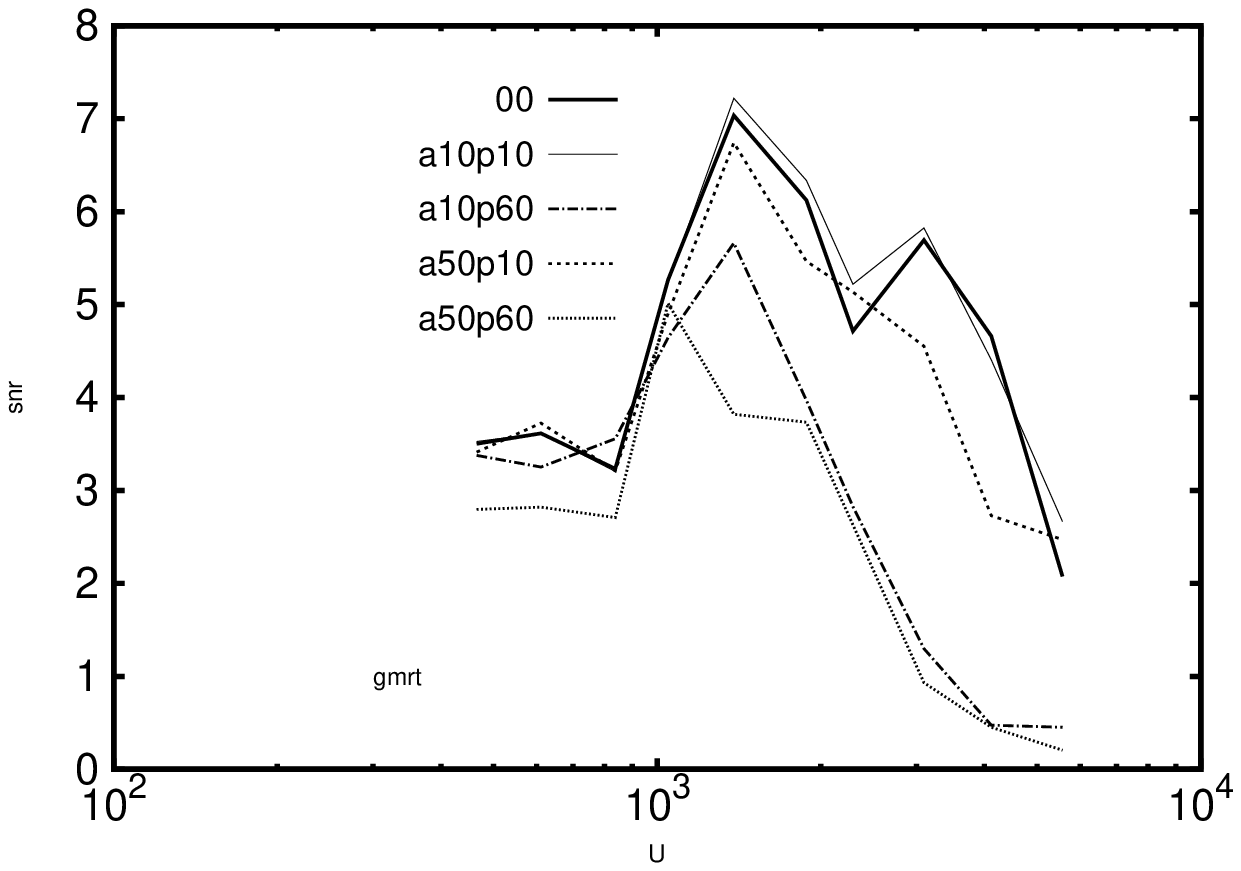}
\caption{The left panel shows the same as 
Figure~\ref{fig:ps1st} for the Tapered Gridded Estimator
 using corrupted visibilities
with the $\sigma_{\alpha}$ and $\sigma_{\phi}$ values shown in the figure.   We have also 
shown $e^{-2\sigma^2_{\phi}} \times C^M_{\ell}$ with $\sigma_{\phi}=60^{\circ}$ for comparison. 
The right panel shows the ${\rm SNR}$ for different values of  $\sigma_{\alpha}$ and $\sigma_{\phi}$.}
\label{fig:gerr}
\end{center}
\end{figure}

We have carried out simulations to test the effect of gain errors on the angular power spectrum
estimators. For this we have generated $20$ different realizations of the random gain errors  and 
used these to corrupt the simulated  visibilities described in Section~\ref{ps_simu}. The  simulations were
carried out for different values of $\sigma_{\alpha}$ and $\sigma_{\phi}$. We have applied both the Bare and the 
Tapered Gridded Estimators on  the corrupted visibilities. Both the estimators show very similar behaviour 
under gain errors, and we show the results for only the Tapered Gridded Estimator.  

 We have considered two values  $\sigma_{\alpha}=0.1$ and $0.5$ which respectively  correspond  to $10 \%$ and
 $50 \%$ errors  in the gain amplitude. The left panel of  Figure~\ref{fig:gerr} shows the results for 
 $\sigma_{\alpha}=0.5$.  We see  that the expectation value of the estimator 
is unaffected by the errors in the  gain amplitude. For the phase errors, we have considered the values
$\sigma_{\phi}=10^{\circ}$ and   $60^{\circ}$ for which $e^{-2\sigma^2_{\phi}}$ have values $0.94$ and $0.11$ 
respectively. The left panel of  Figure~\ref{fig:gerr} shows   that  eq.~(\ref{eq:gerr1}) provides a good
 description for the effect of the gain errors on the 
angular power spectrum estimator. We see the net result of the phase errors  is that the estimated  angular power
spectrum  is reduced by a factor   $e^{-2\sigma^2_{\phi}}$ relative to the input model.

 The right  panel of  Figure~\ref{fig:gerr} shows the ${\rm SNR}$ for 
 different values of  $\sigma_{\alpha}$ and $\sigma_{\phi}$.  The rms. fluctuation  $\sigma_{E_G}$ of  the
 estimator  is expected to  depend exponentially  as $e^{-2\sigma^2_{\phi}}$ on the phase errors and have 
a  $(1+\sigma_{\alpha}^2)^2$ dependence on the amplitude errors (eq.~\ref{eq:vcorrm}). We find that the 
simulated ${\rm SNR}$ are more sensitive to  the phase errors in comparison to the amplitude errors. 
 The   ${\rm SNR}$ is  nearly invariant  to  gain errors
in the cosmic variance dominated regime (low $\ell$) where $\sigma_{E_G}$ 
is  reduced by the same factor $e^{-2\sigma^2_{\phi}}$ as the expectation value of the estimator. 
However, the transition from the cosmic variance dominated to the system noise dominated regime 
(approximately the peak of the  ${\rm SNR}$ curves) shifts to smaller $\ell$ if the gain errors are 
increased.  The amplitude errors, we see, reduces the  ${\rm SNR}$ at large $\ell$ where the error is 
dominated by the system noise.

\section{The  $w$-term}
The entire analysis, until now, has been  based on the assumption that 
the visibility contribution $\S(\u)$ from  the sky signal is  the
Fourier transform of the product  of ${\mathcal A}(\th)$ and $\delta I(\th)$.
This is only an approximate relation which is valid   only if the filed of 
view is sufficiently small. The actual relation is 
\begin{align}
\S(u,v,w)=  \int \, dl dm \frac{\delta I(l,m) {\mathcal A}(l,m)}{\sqrt{1-l^2-m^2}}
e^{-2\pi i[ul+vm+w(\sqrt{1-l^2-m^2}-1)]}\,\,,
\label{eq:weq}
\end{align}
where the $w$-term, which we have ignored until now, is the  baseline component along the line of sight to the 
phase 
center and $l,m$ are the direction cosines corresponding to any point  on the sky.  
In a situation where the primary beam pattern falls of within a small angle from the phase center, 
it is adequate to treat the region of sky under observation as a 2D plane and use $(l,m)=(\theta_x,\theta_y)$. 
For example,  the GMRT  has a FWHM of $186^{'}$   for which  $\sqrt{1-l^2-m^2} \approx 0.997$. The term 
$\sqrt{1-l^2-m^2}$ which appears in the denominator of  eq.~(\ref{eq:weq}) incorporates the curvature 
of the sky.   We see that  this makes an insignificant  contribution at the small angles of our interest, 
and hence  may be ignored. The term $w(\sqrt{1-l^2-m^2}-1)$ which appears in the phase in eq.~(\ref{eq:weq}) 
has a value $\sim 10^{-3} \times w$ for the angle mentioned earlier, and this is not  necessarily small.  
The value of $w$ depends on the telescope configuration and the observing direction, and may be quite large 
$(> 10^3)$.  It is therefore necessary to assess the impact of the $w$-term on the angular power spectrum 
estimators defined earlier.

\begin{figure} 
\begin{center}
\psfrag{u}[t][c][1.3][0]{$\ell$}
\psfrag{err}[b][c][1.3][0]{relative error}
\psfrag{sigma}[r][r][1.][0]{$0.1 \times$ statistical}
\psfrag{1stw}[r][r][1.][0]{$w$-term}
\includegraphics[width=90mm,angle=0]{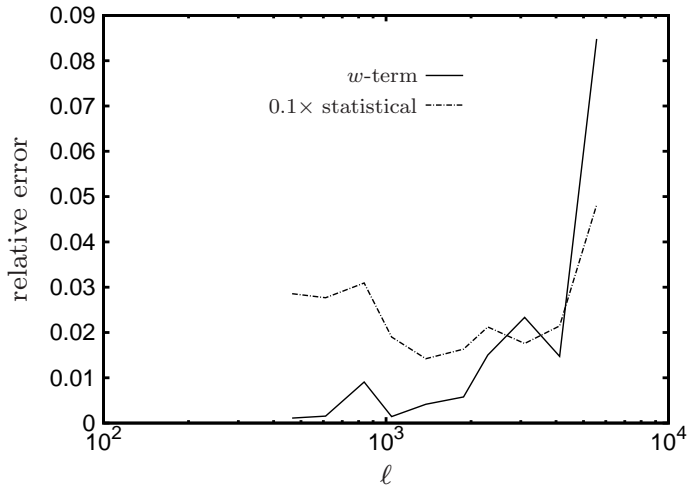}
\caption{This  shows the relative change in the estimated angular power spectrum using Tapered Gridded Estimator due to the $w$-term.  For comparison we have also shown $0.1 \times \delta C_{\ell}/C_{\ell}$ which corresponds to $10 \%$ 
of the relative statistical error in $C_{\ell}$. }
\label{fig:wterm}
\end{center}
\end{figure}

We have simulated  GMRT visibilities  using  eq.~(\ref{eq:weq}) keeping the $w$-term.  The $20$ realizations 
of the sky signal and the baseline tracks are the same as described in section~\ref{ps_simu}, and we have used 
the flat sky approximation ({\it i.e.} we have dropped $\sqrt{1-l^2-m^2}$ from the denominator).  
We have applied both the Bare and the Tapered Gridded Estimator to this simulated visibility data. 
We show results  for only the  Tapered Gridded Estimator, the results are very similar for the  Bare Estimator
and we have not shown these separately.   Figure~\ref{fig:wterm} shows the relative change in the estimated 
angular power spectrum if we include the $w$-term. We find that  the  change due to the $w$-term is less than
 $3 \%$ for all values of $\ell$ barring the largest $\ell$ value where there is a $~9 \%$ change. 
The  $w$-term has a larger effect at the large baselines which also correspond to a larger value of 
$w$.  We find that the change caused by the $w$-term is less than $10 \%$ of the statistical fluctuations 
for most values of $\ell$. In summary, for angular power spectrum estimation  it is adequate to ignore the 
$w$-term at the angular scales of our  interest for the GMRT.

\section{LOFAR}
\label{lofar}
LOFAR, the Low Frequency Array, is an  innovative  new radio
telescope which operates at the lowest radio  frequencies ($10-240
\, {\rm MHz}$) \citep{haarlem}. It
consists of an interferometric array of dipole antenna stations
distributed throughout the Netherlands and Europe. The individual
stations perform the same basic functions as the dishes of a
traditional interferometric radio telescope. Hence, the station beam
which is analogous to the primary beam ultimately determines the FoV
for a given observation. In the High Band Antennas (HBAs, $110 - 240 \,{\rm MHz}$), groups of dipole
pairs are combined into HBA tiles and the station beam is formed from
the combined signal from the tiles. The HBA tiles are sensitive to two orthogonal 
linear polarizations. Close to the phase centre, the
LOFAR station beam can be well modelled with a circular Gaussian and the FWHM  of the Gaussian varies
approximately from $3.0^{\circ}$  to  $5.0^{\circ}$ in the frequency range  $115 \, - 185 \, {\rm MHz}$ 
with  $\theta_{\rm FWHM}=3.8^{\circ}$ at $150 \,  {\rm MHz}$. 

In this section we consider the possibility of using LOFAR to estimate the angular power spectrum 
of the  $150 \,  {\rm MHz}$ sky signal after point source subtraction. 
The LOFAR has a wider field of view compared to the GMRT and 
we have simulated a $\sim \, 8^{\circ}\times 8^{\circ}$ region of the sky 
with an  angular resolution of $14^{''} \times 14^{''}$.  Here again we have 
generated $20$ independent realizations of the sky signal.  The simulations 
were carried out in exactly the same way as described in Section~\ref{ps_simu} 
using the LOFAR parameters given in Table~\ref{tab:1}. 
We have generated the  LOFAR baseline distribution for the 62 
antennas   in the central core region for  $8$ hrs of observing time. 
 Visibilities were generated with a time  interval
of $40$s and  we obtain a total of $669,809$ visibilities 
in the  baseline range $30\le \u\le800$. We have included 
the $w$-term for calculating the LOFAR visibilities. 
 The LOFAR has a denser 
$uv$ coverage compared to the GMRT, and the simulated baseline 
range is nearly uniformly covered.   We have used 
$\sigma_n=2.2$Jy {\citep{haarlem}} for the system  noise   in the simulations. 
Given the large volume of data, we have only used the Tapered Gridded 
Estimator with $f=0.8$ and $w_g=K_{1g}^2$.

\begin{figure}
\begin{center}
\psfrag{cl}[b][b][1.2][0]{$\ell (\ell+1) C_{\ell}/2 \pi \, [mK^2]$}
\psfrag{U}[c][c][1.2][0]{$\ell$}
\psfrag{model}[cr][tr][1.][0]{$C^M_{\ell}$}
\psfrag{lofar}[r][r][1.][0]{LOFAR}
\psfrag{fwt}[r][r][1.][0]{$f=0.8,w_g=\mid K_{1g}\mid^2$}
\includegraphics[width=95mm,angle=0]{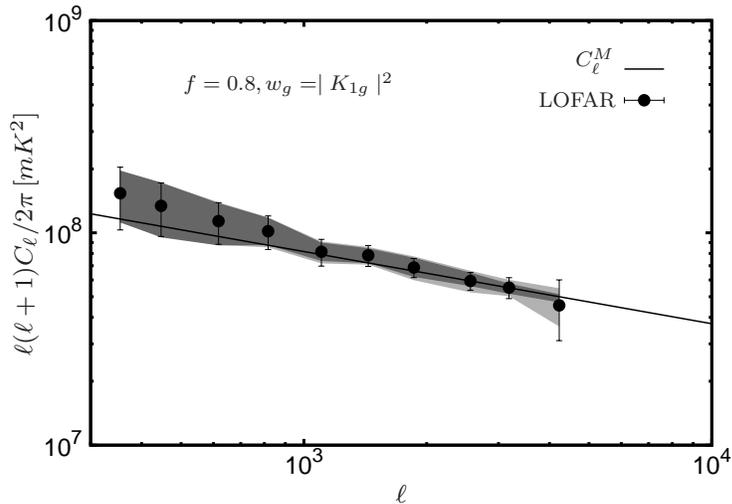}
\caption{Same as Figure~\ref{fig:ps1st} for the Tapered Gridded Estimator and the 
simulated LOFAR data.}
\label{fig:ps2ndlofar}
\end{center}
\end{figure}

Figure  \ref{fig:ps2ndlofar}  shows the angular power spectrum estimated from our simulations. 
We see that  the estimated $C_{\ell}$ values are all  within the $1\sigma$ region of the input model angular
power spectrum $C^{M}_{\ell}$. The estimated $C_{\ell}$ values, however,  are somewhat in excess of 
 $C^M_{\ell}$ at small $\ell$ $(< 1,000)$. The fractional deviation  
$(C_{\ell}-C^M_{\ell})/C^M_{\ell}$ is around $\sim 30 \%$ at the smallest $\ell$ bin, and 
it is  $\sim 15 \%$ at $\ell\sim 800$. 
The excess  is not seen   at larger  $\ell$ 
 where the estimated values are in excellent agreement with  $C^M_{\ell}$.
We also see that  the error estimates predicted by
eq.~(\ref{eq:ge14}) are in good agreement with the rms. fluctuation
estimated from the 20 realizations. The transition from cosmic variance dominated 
errors to  system noise dominated errors occurs at $\ell \sim 2,000$ similar to the GMRT. 
The LOFAR has considerably more baselines compared to the GMRT, and 
the  errors in the estimated  angular power spectrum   are  smaller for LOFAR 
in comparison to GMRT.  

As mentioned earlier in the context of the GMRT, the excess in the estimated
 $C_{\ell}$ may be a consequence of  patchy $uv$ coverage at   
 small baselines  ($U < 160$). The   average baseline density in the region  $U < 160$
is several times larger than the average within  $U < 800$, however this 
does not guaranty that the former is less patchier  than the latter. Further, it
is not possible to say anything definite from a visual inspection of the 
baseline distribution. The convolution with the primary beam pattern and 
the window function introduces a $\sim 8 \%$  deviation between $C_{\ell}$ and 
$C_{\ell}^M$  at $U < 160$. The exact cause of the excess at small $\ell$
is at present not fully understood.

\section{Discussion  and conclusions}
\label{sum}
 In this paper  we have introduced two estimators for quantifying the angular power spectrum of the sky 
brightness temperature. Both of these estimators use the visibilities measured in radio
interferometric observations. The Bare Estimator works directly with the measured visibilities, and 
uses pairwise visibility correlations to estimate the angular power spectrum.  The Tapered Gridded Estimator 
uses the visibility data after gridding on a rectangular grid in the $uv$ plane. Here it is possible to 
taper the sky response so as to suppress the sidelobes and reduce the filed of view.  Earlier work 
\citep{ghosh2} shows tapering  to be an important ingredient in  foreground removal for detecting the 
cosmological $21$-cm signal. We have investigated the properties of the estimators, and present analytic 
formulae for the expectation value (eqs.~\ref{eq:be3} and \ref{eq:ge11a}) and the 
variance (eqs.~\ref{eq:be6} and \ref{eq:ge14}).  The expectation value of both the estimators is free
from the positive system noise bias which arises due to  the correlation of a visibility with itself. The 
system noise  affects only the variance. 

We have carried out simulations to validate the estimators. The simulated sky signal  assumes that the 
point sources have been removed and the residuals are dominated by the diffuse Galactic synchrotron 
radiation which is modelled as   a homogeneous and isotropic Gaussian random field with a power law
 angular power spectrum. We consider GMRT observations for most of the analysis. We find that the Bare 
Estimator is able to recover the input model to a good level of precision.  The computation time is 
found to scale as $N^2$ with the number of visibility data. Further, the scaling is $N^4$ for the variance. 

We find that the Tapered Gridded Estimator is able to recover 
the input model $C_{\ell}^M$ to a high level of precision provided the baselines have a uniform $uv$ coverage. For the GMRT which has a patchy $uv$ coverage,  
 the $C_{\ell}$ estimated from the Tapered Gridded Estimator is largely within the $1\sigma$ errors from  the 
input model $C^M_{\ell}$.  There is, however, indication that  the angular power spectrum 
is  overestimated to some extent. Comparing the results to a situation with  a uniform 
random baseline distribution, 
we conclude that the overestimate is a consequence of GMRT's patchy $uv$ coverage and is not inherent 
to the  Tapered Gridded Estimator  which is unbiased in the ideal situation of  uniform $uv$ coverage.
 It is possible to use simulations to quantify this overestimate and 
correct for  this in a real observation.  We do not anticipate this overestimate to be a very 
major obstacle  for the Tapered Gridded  Estimator. The computation time for this estimator and its 
variance both scale as $N$. Long observations spanning many frequency channels will produce large volumes of
visibility data. The Bare Estimator is computationally very expensive for large $N$, and  a 
Gridded  Estimator is the only feasible alternative. Consequently, we have focused on the Tapered Gridded 
Estimator  for much of the analysis in the later part of this paper. 

Residual gain errors corrupt the measured visibilities, and this is a potential 
difficulty for estimating the angular power spectrum. We have analyzed the effect
of gain errors on the two estimators introduced in this paper. Our analysis, 
validated by simulations, shows that the expectation value of the estimators is 
unaffected by  amplitude errors. The phase errors cause a decrement by the factor 
$e^{-2 \sigma^2_{\phi}}$ in  the expectation value. The statistical errors in the estimated
$C_{\ell}$ are affected by both the amplitude and the phase errors, however this is more 
sensitive to the phase errors relative to the amplitude errors. We have also investigated 
the effect of the $w$-term. We find that the $w$-term does not cause a very big change in the 
estimated $C_{\ell}$ at the scales of our interest here.  Our analysis here shows that 
the residual phase errors can lead to the angular power spectrum being underestimated 
by a factor $e^{-2 \sigma^2_{\phi}}$ which has  a value $\sim 0.1$ for $ \sigma_{\phi}=60^{\circ}$. 
It is therefore  imperative to independently quantify the magnitude of the residual 
phase errors for a correct estimate of the angular power spectrum. 

In addition to GMRT, we have also applied the estimators to  simulated LOFAR data. 
We find that the $C_{\ell}$ estimated using the Tapered Gridded Estimator is within 
the $1\sigma$ errors of the input model. There is, however, indication that there is 
some overestimation  $(15 - 30 \%)$  at low $\ell$ $(< 1,000)$.   The exact cause of this excess at small $\ell$ is at present not fully understood.

The two estimators considered here both avoid  the positive noise bias  which 
arises due to the system noise contribution in the visibilities. This is 
achieved by not including the contribution from the correlation of a visibility
with itself.  As an  alternative one could  consider an estimator which 
straight away squared the measured or the gridded visibilities. In this 
situation it is necessary  to separately identify the noise bias contribution 
and subtract it out. The noise bias contribution is expected to be independent  
of frequency  and $\ell$. It is, in principle, possible to 
identify a  frequency and $\ell$ independent component and subtract it out. 
However, our analysis in this paper shows that the errors in the amplitude 
of the calibrated gains  affect the noise bias. Frequency and baseline 
dependent gain errors would manifest themselves as the frequency and $\ell$ 
dependence of the noise bias. This is a major obstacle which is bypassed
by our estimators. 

The multi-frequency angular power spectrum (MAPS, \citealt{kanan}) 
jointly quantifies the angular and frequency dependence of the fluctuations 
in the  sky signal.   This can be estimated directly from the measured
visibilities (e.g. \citealt{ali}), and it can be  used to  detect the 
cosmological $21$-cm signal \citep{ghosh2}.  In  future work we plan to
generalize the  analysis of this paper to the multi-frequency angular
power spectrum and address various issues, including point source removal, 
which are  relevant  for  detecting the cosmological $21$-cm signal. 

\section{Acknowledgements}
 SC would like to acknowledge UGC, India for providing financial support. SB would like to thank Ayesha Begum, Jayaram N. Chengalur, Prasun Dutta and Jasjeet S. Bagla for useful discussions. AG acknowledge the financial support from the European Research Council under ERC-Starting Grant FIRSTLIGHT - 258942. SSA would like to acknowledge C.T.S, I.I.T. Kharagpur for the use of its facilities and the support by DST, India, under Project No. SR/FTP/PS-088/2010. SSA would also like to thank the authorities of the IUCAA, Pune, India for providing the Visiting Associateship programme.

\end{document}